# Probing the core of the strong nuclear interaction

A. Schmidt et al. (CLAS Collaboration)

**The strong nuclear interaction between nucleons (protons and neutrons) is the effective force that holds the atomic nucleus together. This force stems from fundamental interactions between quarks and gluons (the constituents of nucleons) that are described by the equations of Quantum Chromodynamics (QCD). However, as these equations cannot be solved directly, physicists resort to describing nuclear interactions using effective models that are well constrained at typical inter-nucleon distances in nuclei [1-5] but not at shorter distances. This limits our ability to describe high-density nuclear matter such as in the cores of neutron stars [6]. Here we use high-energy electron scattering measurements that isolate nucleon pairs in short-distance, high-momentum configurations [7-9] thereby accessing a kinematical regime that has not been previously explored by experiments, corresponding to relative momenta above 400 MeV/*c*. As the relative momentum between two nucleons increases and their separation thereby decreases, we observe a transition from a spin-dependent tensor-force to a predominantly spin-independent scalar-force. These results demonstrate the power of using such measurements to study the nuclear interaction at short-distances and also support the use of point-like nucleons with two- and three-body effective interactions to describe nuclear systems up to densities several times higher than the central density of atomic nuclei.**

The binding of nucleons in nuclei disrupts the relationship between their mass ($m_N$), energy ($\epsilon$) and momentum ($p$) such that $\epsilon^2 \neq (m_N c^2)^2 + (pc)^2$. Therefore, describing atomic nuclei requires modeling the interactions of 'off-shell' nucleon pairs and triplets. Modern models of the nuclear interaction, however, are primarily constrained by free (on-shell) nucleon-nucleon (*NN*) scattering data.

Occasionally nucleon pairs at short-distance interact strongly, leading to very high momentum and correspondingly large off-shellness. These naturally occurring high-density fluctuations are called Short-Ranged Correlations (SRCs) [7-9]. We endeavour to test whether modern *NN* interaction models can effectively be used to describe the interaction of these highly off-shell nucleons.

To this end, we measured large momentum-transfer electron scattering from a range of nuclei, studying events where the electron scatters quasielastically (QE) from a bound nucleon, with either one or two protons detected in coincidence with the scattered electron, written as A(*e,e'p*) and A(*e,e'pp*) respectively (see Fig. 1). These measurements are done in kinematical conditions dominated by the hard breakup of SRC pairs.

Our main observation is that in all measured nuclei, from $^{12}$C to $^{208}$Pb, the extracted fraction of *pp*-SRC pairs increases linearly from nucleon momenta of about 400 to about 650 MeV/*c*, and then appears to level off. This indicates a transition from a spin-dependent (tensor) to a spin-independent (scalar) *NN* interaction at high-momenta. This transition is also observed in calculations using either phenomenological or Chiral Effective Field Theory-($\chi$EFT-) based *NN* interaction models, provided that they include a tensor interaction.

The good agreement of the calculations with our data confirms the scalar nature of the *NN* interaction at very high-momenta and validates the use of point-like nucleons with effective interactions for modeling the nuclear interaction. This holds true even where the *NN* interaction is not directly constrained because the nucleons are highly off-shell.

### Electron Scattering SRC Studies

Electron scattering is well described by single-photon exchange [7-14], where electrons scatter from the nucleus by transferring a single virtual photon carrying momentum $\boldsymbol{q}$ and energy ω. In the high-resolution one-body view of QE scattering at large momentum transfer, this virtual photon is absorbed by a single off-shell nucleon with initial energy $\epsilon_i$ and momentum $\boldsymbol{p}_i$.

If the nucleon does not re-interact as it leaves the nucleus, it will emerge with momentum $\boldsymbol{p}_N = \boldsymbol{p}_i + \boldsymbol{q}$ and energy $\epsilon_N = \omega + \epsilon_i$. Outgoing-nucleon rescattering from other nucleons can change the detected momentum and energy. However, we can still approximate the initial momentum and energy of that nucleon as the measured missing momentum $\boldsymbol{p}_{miss} \equiv \boldsymbol{p}_N - \boldsymbol{q} \approx \boldsymbol{p}_i$ and missing energy $E_{miss} \equiv \omega - T_N \approx m_N - \epsilon_i$ (where $T_N = \epsilon_N - m_N$ is the detected-nucleon kinetic energy).

Unlike nucleons in SRC pairs, almost all non-SRC nucleons in atomic nuclei occupy momentum states up to the nuclear Fermi momentum $k_F$ (~ 250 MeV/$c$). Therefore, when $p_{miss} > k_F$, the knockout nucleon should predominantly originate from an SRC pair and should be accompanied by the simultaneous emission of the other (recoil) nucleon with momentum $\boldsymbol{p}_{recoil} \approx -\boldsymbol{p}_i$ [10-13, 15, 16], see Extended Data Fig. 1.

Previous A($e,e'p$) studies observed that more complicated (non-QE) reaction mechanisms can lead to high $p_{miss}$ events that are not due to the knockout of nucleons from SRC pairs. To minimize such contributions, our measurement was performed at kinematics where these non-SRC contributions were shown to be suppressed [8, 17-19], namely: $Q^2 \equiv \boldsymbol{q}^2 - \omega^2 \gtrsim 1.5$ (GeV/$c$)$^2$ and $x_B \equiv Q^2/2\mathrm{m_N}\omega \geq 1.2$, so that $\boldsymbol{p}_{miss}$ was almost anti-parallel to $\boldsymbol{q}$, and $Q^2$ grows with $\boldsymbol{p}_{miss}$. See Methods for details.

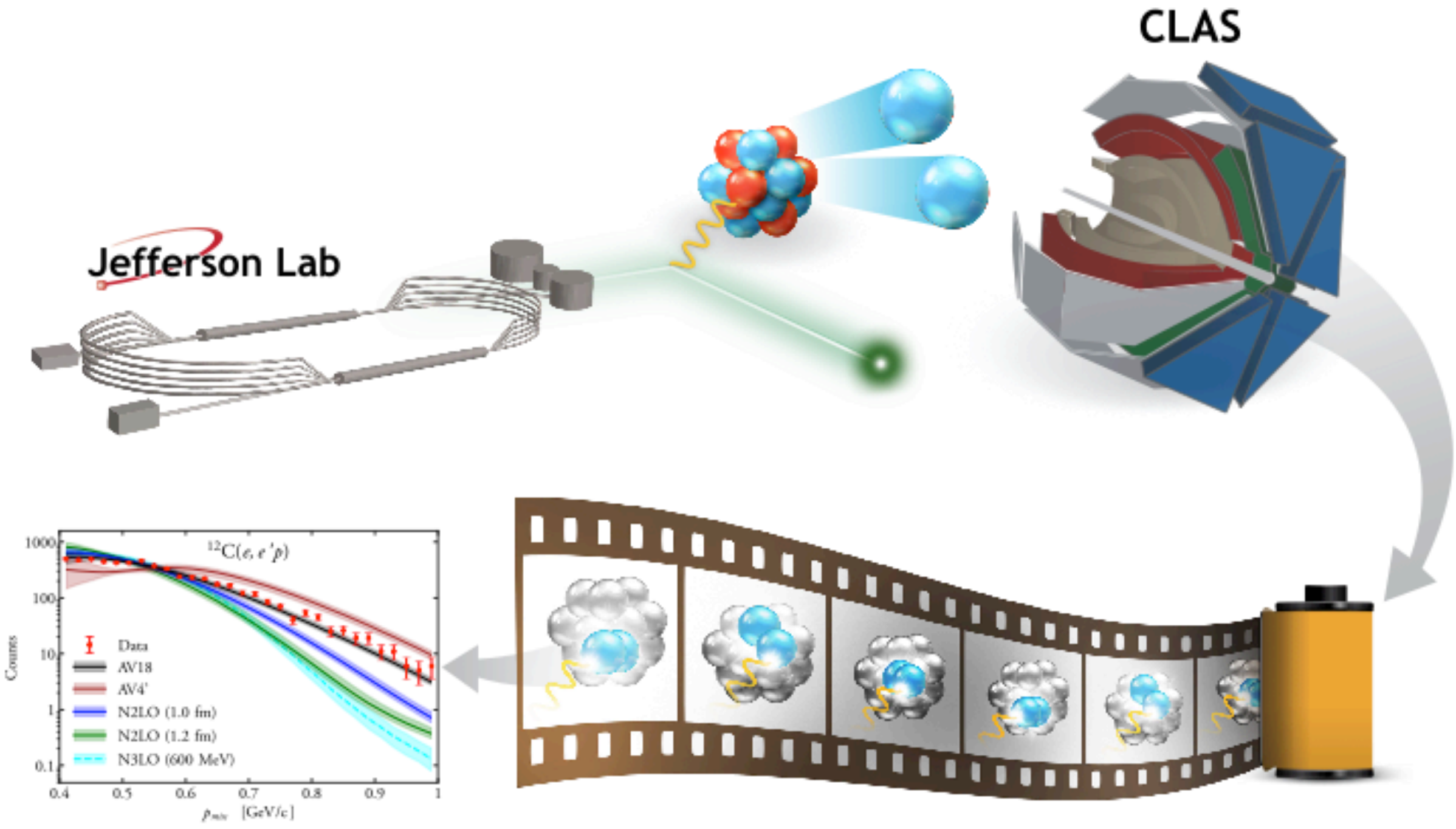

**Fig. 1 | Using Electron Scattering Measurements to Test the Nuclear Interaction.** 5 Giga Electron-Volt (GeV) electrons from the Jefferson Lab accelerator impinge on nuclei and break apart short-range correlated (SRC) nucleon pairs. The CLAS spectrometer is used to detect the scattered electron and knockout protons which allows reconstructing their initial state inside the nucleus. By combining many such measurements, the distribution of such pairs inside the nucleus is assembled and compared to theoretical calculations using different models of the strong nuclear interaction.

Previous measurements of A(*e,e'pN*) reactions off $^4$He and $^{12}$C, performed at similar kinematics, have shown that proton-neutron (*pn*) SRC pairs predominate over proton-proton (*pp*) SRC pairs for $300 < p_{miss} < 600$ MeV/*c* by a factor of almost 20 [10-13, 15]. This is due to the dominance of the tensor part of the *NN* interaction in this momentum range. The tensor force only operates on spin-1 *NN* pairs. As spin-1 *pp*-SRC pairs are suppressed by the Pauli exclusion principle, there are far more *pn*- than *pp*-SRC pairs [7, 8, 17].

At higher missing-momentum where the repulsive core of the *NN* interaction is expected to become dominant, the interaction should be predominantly scalar; i.e. one that operates on both spin-0 and spin-1 pairs. This transition should therefore lead to an increased fraction of *pp*-SRC pairs. Previous work [11] saw initial evidence for such an increase, but its data were statistically limited.

Here we extend these studies by measuring the A(*e,e'p*) and A(*e,e'pp*) reactions for $400 \leq p_{miss} \leq 1000$ MeV/*c* for C, Al, Fe and Pb nuclei. The measurements were performed at the Thomas Jefferson National Accelerator Facility using a 5.01 GeV electron beam. The CEBAF Large Acceptance Spectrometer, CLAS, (Fig. 1) [20] was used to detect and identify the scattered electron and knockout protons and reconstruct their momenta, see Methods for details.

We selected (*e,e'p*) events by considering all measured events with a scattered electron with $x_B \geq 1.2$ and a "leading" proton detected within a narrow cone of 25° around $\boldsymbol{q}$, carrying at least 60% of the transferred momentum ($p_N/q > 0.6$), and resulting in $400 < p_{miss} < 1000$ MeV/*c*. (*e,e'pp*) events are a subset of (*e,e'p*) events where a second, "recoil", proton was detected with momentum greater than 350 MeV/*c*. This recoil proton has significantly smaller momenta and a much wider angular distribution than the high-momentum leading proton. See Extended Data Figs. 2 - 5 for selected kinematical distributions of the measured (*e,e'p*) and (*e,e'pp*) events.

## Cross-Section Modeling

To quantitatively relate observations to the underlying nuclear interaction, we need to calculate the nucleon knockout cross section starting directly from the *NN* interaction.

At the high-$Q^2$ kinematics of our measurement the differential A(*e,e'p*) nucleon knockout cross sections can be approximately factorized as [14, 21]:

Eq. 1 $$\frac{d^6\sigma}{d\Omega_{k'} d\epsilon'_k d\Omega_{p_N} d\epsilon_N} = p_N \epsilon_N \cdot \sigma_{ep} \cdot S(\boldsymbol{p}_i, \epsilon_i),$$

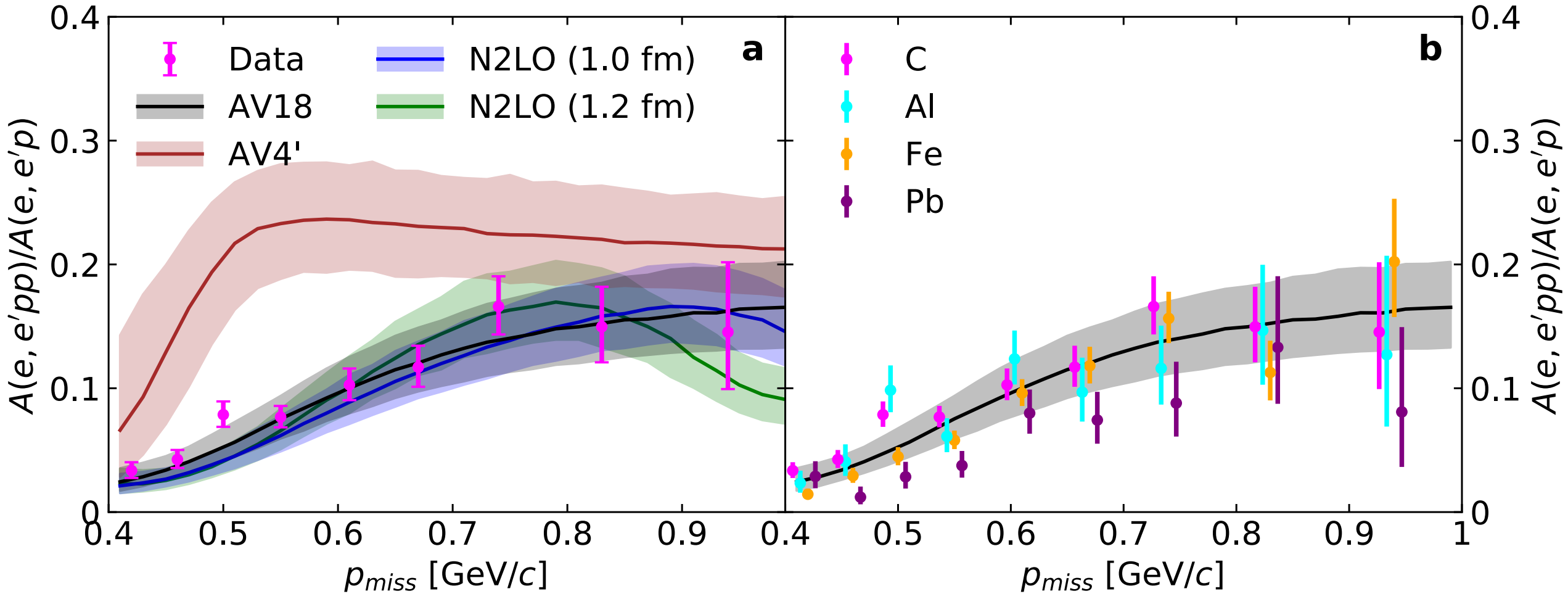

**Fig. 2 | Missing momentum dependence of the two- to one-proton knockout reaction yield ratio.** Measured (*e,e'pp*) / (*e,e'p*) event yields ratios shown as a function of the (*e,e'p*) missing momentum. (a) $^{12}$C data compared with theoretical calculations based on the GCF framework using different models of the *NN* interaction. (b) Comparison of C, Al, Fe, and Pb data and the GCF AV18 $^{12}$C calculation. The width of the bands and the data error bars show the model systematic uncertainties and data statistical uncertainties, respectively, each at the 1σ or 68% confidence level.

where $\boldsymbol{k}'$ and $\epsilon'_k$ are the final electron momentum and energy, $\sigma_{ep}$ is the off-shell electron-nucleon cross section [21] and $S(\boldsymbol{p}_i, \epsilon_i)$ is the nuclear spectral function that defines the probability for finding a nucleon in the nucleus with momentum $\boldsymbol{p}_i$ and energy $\epsilon_i$ [22]. Different models of the nuclear interaction can produce different spectral functions, making the measured cross sections sensitive to the nuclear interaction model.

This commonly used approximation only considers the case where the virtual photon couples to a single nucleon, described using a one-body reaction operator. The full cross-section however also has contributions from many-body operators where the virtual photon couples to more than one nucleon. The contribution of the latter depends on the *NN* interaction model used in the calculation and is very hard to calculate. However, comparisons between experimental data and this model can indicate the size of the many-body contributions in different kinematical regimes and these can later be quantified by more detailed calculations.

The two-nucleon knockout cross section can be factorized similarly to Eq. 1 by replacing the single-nucleon spectral function with the two-nucleon decay function that defines the probability of finding nucleons with momenta $\boldsymbol{p}_i$ and $\boldsymbol{p}_{recoil}$ such that the A-1 system (the A-2 nucleus plus the recoil proton) has energy $E_r$ [9, 15, 17]. See Supplementary Information for details.

Ab-initio many-body calculations of the nuclear spectral and decay functions are currently computationally unfeasible [1]. However, for the specific case of interacting with SRC pairs (i.e. $p_i \approx p_{miss} > k_F$), we can effectively approximate these functions using the Generalized Contact Formalism (GCF) [22-25] which assumes that at very high momenta, the nuclear wave-function can be described as consisting of an SRC pair and a residual A-2 system. The abundance of SRC pairs is given by nuclear contact terms extracted from ab-initio many-body calculations of pair momentum distributions [24, 25].

Therefore, in the GCF, the high-momentum proton spectral function of Eq. 1 is approximated by a sum over *pp* and *pn* SRC pairs, which allows calculating (*e,e'p*) and (*e,e'pp*) cross sections using different nuclear interaction models as input [13, 22].

We consider two commonly used *NN* interaction models: the phenomenological AV18 [4] and the $\chi$EFT local N2LO [5] interactions, as well as the simplified, tensor-less, AV4' interaction. The $\chi$EFT potentials considered here include explicit cutoffs at distances of 1.0 fm and 1.2 fm corresponding to momentum cutoffs of about $400 - 500$ MeV/c [26]. We do not expect these interactions to work well above this cutoff (see Methods for details).

We compared the GCF cross sections to experimental data using Monte Carlo integration, accounting for the CLAS acceptance, resolution, and residual reaction effects (radiation, transparency and single-charge exchange). The systematic uncertainty of the calculation was estimated by varying the GCF and detector model parameters. See Methods for details on the GCF model and its implementation.

## Measurement Results

Fig. 2 shows the measured (*e,e'pp*) / (*e,e'p*) event yield ratio as a function of $p_{miss}$ for C, Al, Fe and Pb. The ratio increases linearly from 400 to about 650 MeV/*c* and then appears to flatten out for all measured nuclei. The observed increase in this ratio, i.e., the fraction of (*e,e'p*) events with a recoil proton, is qualitatively consistent with the expected transition from a predominantly tensor to a predominantly scalar interaction at high $p_{miss}$.

For $^{12}$C, the measured ratio is compared with GCF calculations using the AV18 and N2LO interactions that are in excellent agreement with the data. The scalar-only AV4' interaction (i.e., which lacks the tensor force) agrees with data in the scalar-dominated high-momentum region but fails in the tensor-dominated low-momentum region.

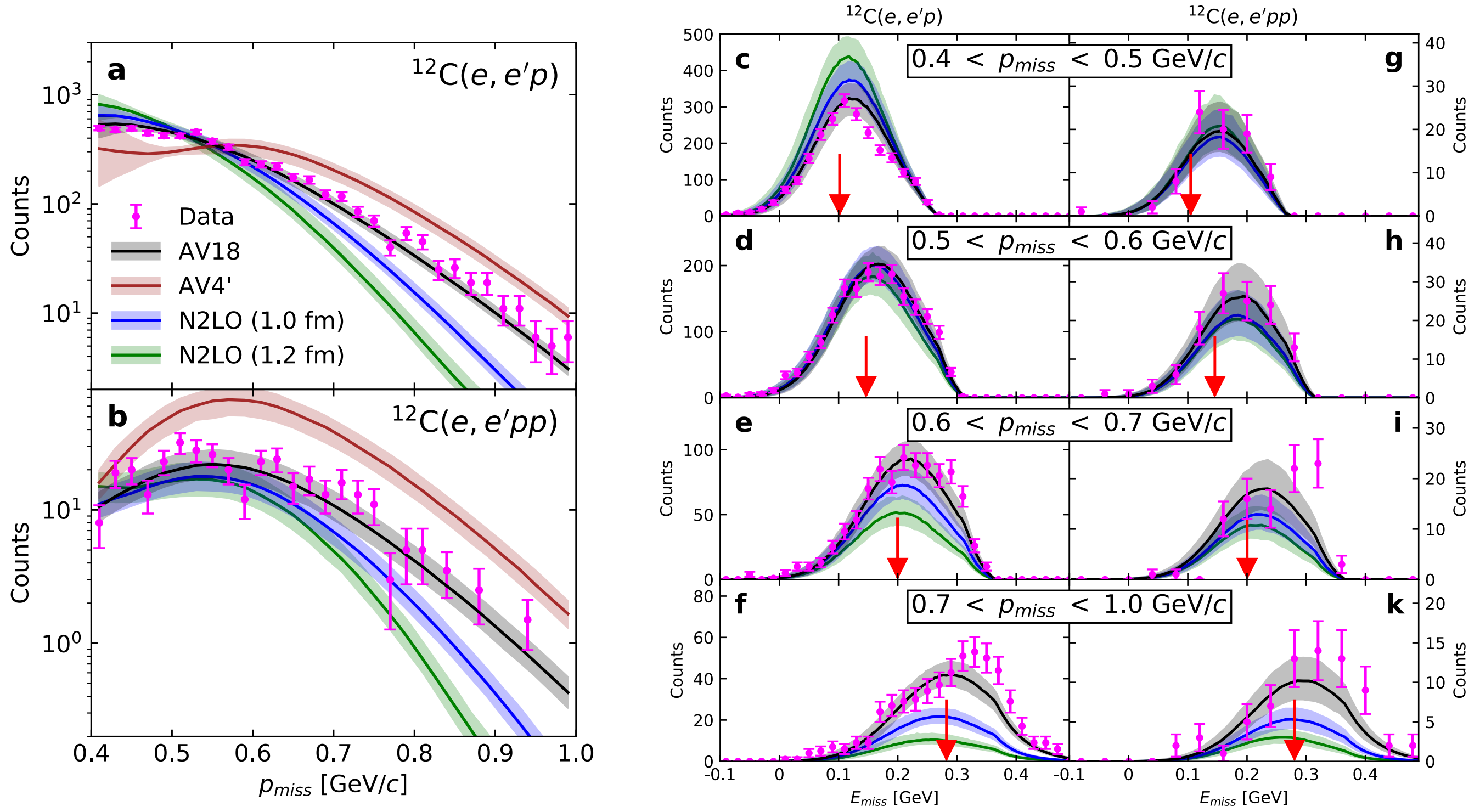

**Fig. 3 | Missing momentum and energy dependence of one- and two-proton knockout reaction yields.** Measured $^{12}$C(*e,e'p*) (a, c-f) and $^{12}$C(*e,e'pp*) (b, g-j) event yields shown as a function of the (*e,e'p*) missing momentum (a,b) and missing energy (c-j). The data are compared with theoretical calculations based on the GCF framework, using different models of the *NN* interaction. The arrows mark the expected missing energy for interacting with a stationary pair with relative momentum that equals the mean of each missing-momentum bin. The width of the bands and the data error bars show the model systematic uncertainties and data statistical uncertainties, respectively, each at the 1σ or 68% confidence level.

At high-momenta all calculations predict a *pp*-SRC pair fraction of ~1/3 (Extended Data Fig. 6 (c)), which is equal to the scalar limit one obtains by simple pair counting (see calculation in Methods). This value of 1/3 is then reduced experimentally by the CLAS acceptance and residual reaction effects.

Fig. 3 shows the absolute measured and calculated $^{12}$C(*e,e'pp*) and $^{12}$C(*e,e'p*) yields as a function of $p_{miss}$ and as a function of $E_{miss}$ for different bins in $p_{miss}$. The average value of $E_{miss}$ increases with $p_{miss}$, peaking at the expected value for the breakup of an SRC pair at rest, marked by a red arrow in Fig. 3 (see Eq. 2 in Methods). This supports our interpretation of the measured process being dominated by interacting with an SRC pair with the A-2 residual system being a spectator [27].

The GCF calculations follow the same trend as the data. The AV18 interaction agrees with the data over the entire $E_{miss}$ and $p_{miss}$ range. The simplified AV4' interaction, as expected, does not describe the momentum distributions well. The N2LO interactions describe the data well up to about 600 MeV/c. The latter observation is surprising since the $\chi$EFT cutoff truncates the *NN* interaction quite severely, leading to very large expected uncertainties already at the cutoff scale. The fact that interactions studied here use position-space regulator makes their truncation effects significant at a high momentum scale of 600 – 700 MeV/c.

While for $p_{miss} > 600$ MeV/c the $\chi$EFT calculations disagree with the individual $^{12}$C(e,e'pp) and $^{12}$C(e,e'p) yields, as expected. This disagreement however largely cancels in the (e,e'pp) / (e,e'p) yield ratio, indicating that such ratios are good observables that are sensitive to the operator structure of the NN interaction.

We therefore use the (*e,e'pp*) / (*e,e'p*) ratios to extract the relative abundance of *np*/*pp* SRC pairs (i.e. contact term ratios) by fitting the contacts in the GCF calculation to the data (see Methods for details). The resulting contact ratios are listed in Extended Data Table 1. The C contact ratios are consistent with those extracted from ab-initio calculations [24, 25] and the ratios of heavier nuclei relative to $^{12}$C are observed to be model-independent quantities, as expected [25].

Extended Data Figs. 3 and 4 show good agreement between the data and GCF calculation for various other kinematical distributions.

The agreement between the data and the AV18 interaction-based calculations corroborates the validity of the assumptions made in the GCF model, namely the dominance of the interaction of an SRC nucleon pair with the electron via a one-body current operator. On the other hand, we stress that the disagreement we find at higher momentum between the $\chi$EFT-based calculations and the data does not necessarily indicate a problem with the *NN* interaction. As the theoretical cross section is sensitive to both the nuclear interaction and the current operator, the shortcomings of the calculations might be attributed to two-body currents. Additionally, the approximations made in the GCF model, e.g. neglecting three-body SRCs, may explain some of the disagreement.

Two additional tests confirm the suppression of non-QE reaction mechanisms: (1) the A(*e,e'p*) and A(*e,e'pp*) $p_{miss}$ and $E_{miss}$ distributions for nuclei from C to Pb are identical within uncertainties, indicating the suppression of $A$-dependent non-QE reaction mechanisms, and (2) the distribution of the kinematical variables that are most sensitive to non-QE reaction mechanisms, such as the angle between $\boldsymbol{p}_{miss}$ and $\boldsymbol{q}$ [8, 17, 19], are well described by the GCF-based simulation, see Extended Data Fig. 3 and Supplementary Information for details.

Lastly, due to the high initial-momenta of the measured protons, we assessed the possible impact of relativistic effects on the nuclear wave-function in the GCF spectral function. As fully relativistic nuclear potentials and wave-functions are unavailable, relativistic effects can only be treated in an approximate and model-dependent manner. Here we used the relativistic nuclear light cone (LC) formalism of [9], which was previously used for SRC studies using nucleon knockout reactions [15], see Supplementary Information for details.

Extended data Fig. 7 shows relativistic LC calculations compared with the same data shown in Fig 3. The relativistic corrections somewhat reduce the agreement with (*e,e'pp*) data at lower momenta and significantly improve the agreement of $\chi$EFT-based calculations with both (*e,e'p*) and (*e,e'pp*) data at higher momenta. This is because, in kinematics of our measurement, the relativistic treatment reduces the effective relative momenta of the probed *NN* pairs, bringing it closer to the $\chi$EFT cutoff scale. This suggests that the importance of two-body operators in $\chi$EFT-based calculations at large $p_{miss}$ might be small, and should be studied in the future using dedicated relativistic calculations.

## Conclusions

The measured A(*e,e'pp*) / A(*e,e'p*) cross-section ratio is observed to be nucleus independent, indicating a transition from behavior reflecting the tensor character of *NN* pairs at to behavior described by spin-independent correlations at high $p_{miss}$. The large momentum transfer electron scattering measurements reported here are thus sensitive to the detailed characteristics of the *NN* interaction at high relative momenta.

The one-body GCF approximation describes the data well up to very high missing momenta, even though the input *NN* interaction models were not directly fit to high-momentum data.

These *NN* interactions result from the quark-gluon structure of nucleons. Measurements of quark distributions in nuclei are often explained by modifying bound-nucleon structure [7, 28]. Such modifications were recently associated with the large spatial overlap and large off-shellness of nucleons in SRC pairs [7, 29, 30]. Our results however suggest that even if such modifications exist, they do not significantly impact the effective modeling of the *NN* interaction.

Our data also point to the importance of two-body interaction operators and relativistic effects at high-momenta and calls for more elaborate theoretical calculations. Combined with forthcoming three-nucleon knockout measurements, such calculations will also allow for future studies of the loosely constrained three-nucleon (i.e. *NNN*) interaction.

Lastly, we provide strong support for the use of point-like nucleons with effective interactions for modeling both atomic nuclei and dense astrophysical systems such as neutron stars, whose outer core already exceeds the nuclear saturation density.

**References:**

[1] "Quantum Monte Carlo methods for nuclear physics", J. Carlson, S. Gandolfi, F. Pederiva, S. C. Pieper, R. Schiavilla, K. E. Schmidt, and R. B. Wiringa, Rev. Mod. Phys. **87**, 1067 (2015).

[2] "Modern theory of nuclear forces", E. Epelbaum, H.W. Hammer and U.G. Meißner, Rev. Mod. Phys. **81**, 1773 (2009).

[3] "The bonn meson-exchange model for the nucleon-nucleon interaction", R. Machleidt, K. Holinde, and C. Elster, Phys. Rept. **149**, 1-89 (1987).

[4] "Accurate nucleon-nucleon potential with charge-independence breaking", R. B. Wiringa, V. G. J. Stoks, and R. Schiavilla, Phys. Rev. C **51**, 38 (1995).

[5] "Quantum Monte Carlo Calculations with Chiral Effective Field Theory Interactions", A. Gezerlis, I. Tews, E. Epelbaum, S. Gandolfi, K. Hebeler, A. Nogga, and A. Schwenk, Phys. Rev. Lett. **111**, 032501 (2013).

[6] "Neutron Star Observations: Prognosis for Equation of State Constraints" J.M. Lattimer and M. Prakash, Phys. Rept. **442**, 109-165 (2007).

[7] "Nucleon-Nucleon Correlations, Short-lived Excitations, and the Quarks Within", O. Hen, G.A. Miller, E. Piasetzky, and L. B. Weinstein, Rev. Mod. Phys. **89**, 045002 (2017).

[8] "In-medium short-range dynamics of nucleons: Recent theoretical and experimental advances", C. Ciofi degli Atti, Phys. Rep. **590**, 1-85 (2015).

[9] "High-Energy Phenomena, Short Range Nuclear Structure and QCD", L. Frankfurt, and M. Strikman, Phys. Rep. **76**, 215-347 (1981).

[10] "Probing Cold Dense Nuclear Matter", R. Subedi et al., Science **320**, 1476-1478 (2008).
[11] "Approaching the nucleon-nucleon short-range repulsive core via the 4He(e,e'pN) triple coincidence reaction", I. Korover et al., Phys. Rev. Lett. **113**, 022501 (2014).
[12] "Momentum Sharing in Imbalanced Fermi Systems", O. Hen et al. (CLAS Collaboration), Science **346**, 614-617 (2014).
[13] "Direct Observation of Proton-Neutron Short-Range Correlation Dominance in Heavy Nuclei", M. Duer et al. (CLAS Collaboration), Phys. Rev. Lett. **122**, 172502 (2019).
[14] "Nucleon knockout by intermediate-energy electrons", J.J. Kelly, Adv. Nucl. Phys. **23**, 75-294 (1996).
[15] "Evidence for the strong dominance of proton-neutron correlations in nuclei", E. Piasetzky, M. Sargsian, L. Frankfurt, M. Strikman, and J.W. Watson, Phys. Rev. Lett. **97**, 162504 (2006).
[16] "Center of Mass Motion of Short-Range Correlated Nucleon Pairs studied via the A(e,e'pp) Reaction", E. O. Cohen et al. (CLAS Collaboration), Phys. Rev. Lett. **121**, 092501 (2018).
[17] "Recent observation of short range nucleon correlations in nuclei and their implications for the structure of nuclei and neutron stars", L. Frankfurt, M. M. Sargsian, and M. Strikman, Int. J. Mod. Phys. A **23**, 2991-3055 (2008).
[18] "Final-state interactions in two-nucleon knockout reactions", C. Colle, W. Cosyn, and J. Ryckebusch, Phys. Rev. C **93**, 034608 (2016).
[19] "Selected Topics in High Energy Semi-Exclusive Electro-Nuclear Reactions", M. M. Sargsian, Int. J. Mod. Phys. E **10**, 405-458 (2001).
[20] "The CEBAF large acceptance spectrometer (CLAS)", B. A. Mecking et al., Nucl. Inst. Meth. A **503**, 513-553 (2003).
[21] "Off-shell electron-nucleon cross sections: The impulse approximation", T. De Forest, Nucl. Phys. A **392**, 232-248 (1983).
[22] "Energy and momentum dependence of nuclear short-range correlations - Spectral function, exclusive scattering experiments and the contact formalism", R. Weiss, I. Korover, E. Piasetzky, O. Hen, and N. Barnea, Phys. Lett. B **791**, 242-248 (2019).
[23] "Generalized nuclear contacts and momentum distributions", R. Weiss, B. Bazak, and N. Barnea, Phys. Rev. C **92**, 054311 (2015).
[24] "The Nuclear Contacts and Short-Range Correlations in Nuclei", R. Weiss, R. Cruz-Torres, N. Barnea, E. Piasetzky, and O. Hen, Phys. Lett. B **780**, 211-215 (2018).
[25] "Scale and Scheme Independence and Position-Momentum Equivalence of Nuclear Short-Range Correlations" R. Cruz-Torres, D. Lonardoni, R. Weiss, N. Barnea, D.W. Higinbotham, E. Piasetzky, A. Schmidt, L.B. Weinstein, R.B. Wiringa, and O. Hen, arXiv: 1907.03658 (2019).
[26] "Weinberg eigenvalues for chiral nucleon-nucleon interactions", J. Hoppe, C. Drischler, R. J. Furnstahl, K. Hebeler, and A. Schwenk Phys. Rev. C 96, 054002 (2017).
[27] "Two nucleon correlations and the structure of the nucleon spectral function at high values of momentum and removal energy", C. Ciofi degli Atti, S. Simula, L.L. Frankfurt, and M.I. Strikman, Phys. Rev. C **44**, R7-R11 (1991).
[28] "The EMC Effect", P.R. Norton, Rept. Prog. Phys. **66**, 1253-1297 (2003).
[29] "Global study of nuclear structure functions", S. A. Kulagin and R. Petti, Nucl. Phys. A **765**, 126-187 (2006).
[30] "Modified Structure of Protons and Neutrons in Correlated Pairs", B. Schmookler et al. (CLAS Collaboration), Nature **566**, 354-358 (2019).

**Acknowledgements** We acknowledge the efforts of the staff of the Accelerator and Physics Divisions at Jefferson Lab that made this experiment possible. The analysis presented here was carried out as part of the Jefferson Lab Hall B Data-Mining project supported by the U.S. Department of Energy (DOE). The research was supported also by the National Science Foundation, the Israel Science Foundation, the Pazi foundation, the Chilean Comisión Nacional de Investigación Científica y Tecnológica, the French Centre National de la Recherche Scientifique and Commissariat a l'Energie Atomique, the French-American Cultural Exchange, the Italian Istituto Nazionale di Fisica Nucleare, the National Research Foundation of Korea, and the UKs Science and Technology Facilities Council. Jefferson Science Associates operates the Thomas Jefferson National Accelerator Facility for the DOE, Office of Science, Office of Nuclear Physics under contract DE-AC05-06OR23177.

**Author Contributions** The CEBAF Large Acceptance Spectrometer was designed and constructed by the CLAS Collaboration and Jefferson Lab. Data acquisition, processing and calibration, Monte Carlo simulations of the detector, and data analyses were performed by a large number of CLAS Collaboration members, who also discussed and approved the scientific results. The analysis presented here was performed primarily by A.S. and J.P. R.W. and N.B. provided theoretical input and helped implement parts of the GCF event generator. M.S. and A.L. provided theoretical input and helped implement the Light Cone Formalism. A.D. and E.S. helped implement parts of the GCF event generator and performed the model systematic uncertainty studies. A.H. calculated the CLAS acceptance maps. O.H., E.P., and L.B.W. guided and supervised the analysis.

**Data Availability** The raw data from this experiment are archived in Jefferson Laboratory's mass storage silo.

**The CLAS Collaboration**

A. Schmidt,[1,2] J.R. Pybus,[1] R. Weiss,[3] E.P. Segarra,[1] A. Hrnjic,[1] A. Denniston,[1] O. Hen, [1,+] E. Piasetzky,[4] L.B. Weinstein,[5] N. Barnea,[3] M. Strikman,[6] A. Larionov,[7] D. Higinbotham,[8] S. Adhikari,[9] M. Amaryan,[5] G. Angelini,[2] G. Asryan,[10] H. Atac,[11] H. Avakian,[8] C. Ayerbe Gayoso,[12] L. Baashen,[9] L. Barion,[13] M. Bashkanov,[14] M. Battaglieri,[8, 15] A. Beck,[1] I. Bedlinskiy,[16] F. Benmokhtar,[17] A. Bianconi,[18, 19] A.S. Biselli,[20, 21] F. Bossù,[22] S. Boiarinov,[8] W.J. Briscoe,[2] W. Brooks,[24] V.D. Burkert,[8] F. Cao,[25] D.S. Carman,[8] J.C. Carvajal,[9] A. Celentano,[15] P. Chatagnon,[26] T. Chetry,[27] G. Ciullo,[13, 28] L. Clark,[29] E. Cohen,[4] P.L. Cole,[30, 31, 32] M. Contalbrigo,[13] V. Crede,[33] R. Cruz-Torres,[1] A. D'Angelo,[34, 35] N. Dashyan,[10] R. De Vita,[15] E. De Sanctis,[36] M. Defurne,[22] A. Deur,[8] S. Diehl,[25] C. Djalali,[37, 38] M. Duer,[4] M. Dugger,[39] R. Dupre,[26] H. Egiyan,[8] M. Ehrhart,[26] A. El Alaoui,[24] L. El Fassi,[27] P. Eugenio,[33] A. Filippi,[40] T.A. Forest,[31] G. Gavalian,[8, 41] S. Gilad,[1] G.P. Gilfoyle,[42] K.L. Giovanetti,[43] F.X. Girod,[8, 13] C. Giuseppe,[8, 13] D.I. Glazier,[29] E. Golovatch,[44] R.W. Gothe,[38] K.A. Griffioen,[12] L. Guo,[9] K. Hafidi,[23] H. Hakobyan,[10, 24] C. Hanretty,[8] N. Harrison,[8] M. Hattawy,[5] F. Hauenstein,[1, 5] T.B. Hayward,[12] K. Hicks,[37] M. Holtrop,[41] Y. Ilieva,[2, 38] I. Illari,[2] D. Ireland,[29] B.S. Ishkanov,[44] E.L. Isupov,[44] D. Jenkins,[45] H.S. Jo,[46] K. Joo,[25] D. Keller,[51] M. Khachatryan,[5] A. Khanal,[9] M. Khandaker,[47, *] C.W. Kim,[2] W. Kim,[46] F.J. Klein,[32] I. Korover,[4] V. Kubarovsky,[8, 48] L. Lanza,[34] M. Leali,[18, 19] P. Lenisa,[13] I.J.D. MacGregor,[29] D. Marchand,[26] N. Markov,[25] V. Mascagna,[19, 49, †] S. May-Tal Beck,[1] B. McKinnon,[29] M. Mirazita,[36] V. Mokeev,[8] C. Munoz Camacho,[26] B. Mustapha,[23] P. Nadel-Turonski,[8] S. Nanda,[27] S. Niccolai,[26] G. Niculescu,[43] M. Osipenko,[15] A.I. Ostrovidov,[33] M. Paolone,[11] L.L. Pappalardo,[13] R. Paremuzyan,[41] K. Park,[46, ‡] E. Pasyuk,[8, 39] M. Patsyuk,[1] W. Phelps,[2] O. Pogorelko,[16] J.W. Price,[50] Y. Prok,[5, 51] D. Protopopescu,[29] M. Ripani,[15] D. Riser,[25] A. Rizzo,[34, 35] G. Rosner,[29] P. Rossi,[8, 36] F. Sabatié,[22] C. Salgado,[47] B. Schmookler,[1] R.A. Schumacher,[21] Y.G. Sharabian,[8] U. Shrestha,[37] Iu. Skorodumina,[38, 44] D. Sokhan,[29] O. Soto,[36] N. Sparveris,[11] S. Stepanyan,[8] I.I. Strakovsky,[2] S. Strauch,[2, 38] N. Tyler,[38] M. Ungaro,[8, 48] L. Venturelli,[18, 19] H. Voskanyan,[10] E. Voutier,[26] R. Wang,[26] D.P. Watts,[14] X. Wei,[8] M.H. Wood,[38, 52] N. Zachariou,[14] J. Zhang,[51] Z.W. Zhao,[53] and X. Zheng[51]

[1]Massachusetts Institute of Technology, Cambridge, MA 02139
[2]The George Washington University, Washington, DC 20052
[3]Hebrew University, Jerusalem, Israel
[4]Tel Aviv University, Tel Aviv, Israel
[5]Old Dominion University, Norfolk, Virginia 23529
[6]Pennsylvania State University, University Park, PA 16802
[7]Frankfurt Institute for Advanced Studies, Giersch Science Center, D-60438 Frankfurt am Main, Germany
[8]Thomas Jefferson National Accelerator Facility, Newport News, Virginia 23606
[9]Florida International University, Miami, Florida 33199
[10]Yerevan Physics Institute, 375036 Yerevan, Armenia
[11]Temple University, Philadelphia, PA 19122
[12]College of William and Mary, Williamsburg, Virginia 23187-8795
[13]INFN, Sezione di Ferrara, 44100 Ferrara, Italy
[14]University of York, York YO10 5DD, United Kingdom
[15]INFN, Sezione di Genova, 16146 Genova, Italy
[16]National Research Centre Kurchatov Institute - ITEP, Moscow, 117259, Russia
[17]Duquesne University, Pittsburgh, Pennsylvania. 15282 USA
[18]Universit'a degli Studi di Brescia, 25123 Brescia, Italy
[19]INFN, Sezione di Pavia, 27100 Pavia, Italy
[20]Fairfield University, Fairfield CT 06824
[21]Carnegie Mellon University, Pittsburgh, Pennsylvania 15213
[22]IRFU, CEA, Universit'e Paris-Saclay, F-91191 Gif-sur-Yvette, France
[23]Argonne National Laboratory, Argonne, IL 60439
[24]Universidad T´ecnica Federico Santa Mar´ıa, Casilla 110-V Valpara´ıso, Chile
[25]University of Connecticut, Storrs, Connecticut 06269
[26]Institut de Physique Nucl'eaire, IN2P3-CNRS, Universit'e Paris-Sud, Universit'e Paris-Saclay, F-91406 Orsay, France
[27]Mississippi State University, Mississippi State, MS 39762-5167
[28]Universita' di Ferrara , 44121 Ferrara, Italy
[29]University of Glasgow, Glasgow G12 8QQ, United Kingdom
[30]Lamar University, 4400 MLK Blvd, PO Box 10009, Beaumont, Texas 77710
[31]Idaho State University, Pocatello, Idaho 83209
[32]Catholic University of America, Washington, D.C. 20064

[33]Florida State University, Tallahassee, Florida 32306
[34]INFN, Sezione di Roma Tor Vergata, 00133 Rome, Italy
[35]Universita' di Roma Tor Vergata, 00133 Rome Italy
[36]INFN, Laboratori Nazionali di Frascati, 00044 Frascati, Italy
[37]Ohio University, Athens, Ohio 45701
[38]University of South Carolina, Columbia, South Carolina 29208
[39]Arizona State University, Tempe, Arizona 85287-1504
[40]INFN, Sezione di Torino, 10125 Torino, Italy
[41]University of New Hampshire, Durham, New Hampshire 03824-3568
[42]University of Richmond, Richmond, Virginia 23173
[43]James Madison University, Harrisonburg, Virginia 22807
[44]Skobeltsyn Institute of Nuclear Physics, Lomonosov Moscow State University, 119234 Moscow, Russia
[45]Virginia Tech, Blacksburg, Virginia 24061-0435
[46]Kyungpook National University, Daegu 41566, Republic of Korea
[47]Norfolk State University, Norfolk, Virginia 23504
[48]Rensselaer Polytechnic Institute, Troy, New York 12180-3590
[49]Universita degli Studi dellInsubria, 22100 Como, Italy
[50]California State University, Dominguez Hills, Carson, CA 90747
[51]University of Virginia, Charlottesville, Virginia 22901
[52]Canisius College, Buffalo, NY 14208
[53]Duke University, Durham, North Carolina 27708-0305

[+]Contact Author: hen@mit.edu
[*]Current address: Idaho State University, Pocatello, Idaho 83209
[†]Current address: Universit'a degli Studi di Brescia, 25123 Brescia, Italy
[‡]Current address: Thomas Jefferson National Accelerator Facility, Newport News, Virginia 23606

# Methods:

### CLAS Detector and Particle Identification

CLAS was a 6-sector toroidal magnetic spectrometer [20]. Each sector was equipped with three layers of drift chambers, time-of-flight scintillation counters, Cerenkov counters, and electromagnetic calorimeters. The drift chambers and time-of-flight scintillation counters covered in-plane scattering angles from about 8° to 140°, while the Cerenkov counters and electromagnetic calorimeters covered about 8° to 45°. The 6 sectors collectively covered 50-80% of the out-of-plane angle.

Charged particles' positions were measured in the drift chambers, allowing reconstruction of their trajectories as they bent due to the influence of the toroidal magnetic field. The charge of charged particles (electrons and protons in this work) and their momenta were determined from their reconstructed trajectories. We consider only charged particles whose trajectories were reconstructed to originate in the location of the solid target foil, see [31] for details.

Electrons were distinguished from pions by requiring a large signal in the Cerenkov counters, as well as a large energy deposition in the Electromagnetic Calorimeters that is proportional to momentum. Protons were identified by requiring that their time-of-flight, measured by the scintillation counters, be within two standard deviations of the calculated time-of-flight based on the momentum reconstructed in the drift chambers, assuming the particle has the mass of a proton.

### Measurement Kinematics and Reaction Mechanism Effects:

Experimentally, we measure final-state particles and reconstruct the initial state of the nucleons, before the electron interaction, based on modeling of the electron scattering reaction using the GCF to model the nuclear spectral and decay functions [9, 15, 17, 22-25, 32-34]. This work focuses on the specific interpretation of the data in terms of QE electron scattering from a single nucleon, as shown in Extended Data Fig. 1. However, as shown by previous studies, the reaction can also include contributions from (1) meson-exchange currents (MEC), (2) isobar currents (exciting the struck nucleon to an intermediate excited state), (3) elastic and inelastic nucleon rescattering (final-state interactions, FSIs), and (4) single charge exchange (SCX) reactions, that would all lead to a similar final state as the QE scattering reaction. The relative contribution of these reaction mechanisms depends on the kinematics of the experiment [17-19, 35-39], see Ref. [7, 8] and references therein for a detailed discussion and review of previous experimental and theoretical studies. For example, Isobar currents are suppressed for $x_B > 1$, as, for a given $Q^2$, the virtual photon transfers less energy and is less likely to excite the nucleon to a higher energy state.

For high missing momentum events, elastic FSIs include rescattering of the outgoing nucleon with the other nucleon of the SRC pair or with the other nucleons in the residual nucleus. At large knock-out nucleon momenta, such rescattering, as well as SCX interactions, can be estimated using a generalized Eikonal approximation in a Glauber framework [18, 19, 40], previously shown to well-reproduce experimental data [41-43]. These calculations show that in our kinematics, elastic FSIs are largely confined to nucleons in close proximity, and the largest part of the scattering cross section can be attributed to rescattering between nucleons of the SRC pair [44]. Therefore, FSI predominantly leads to flux reduction that can be quantified in terms of a transparency factor.

SCX can turn proton knockout events into neutron knockout events – reducing the measured proton yield – as well as turn neutron knockout events into proton knockout events – increasing the measured proton yield. The rate of both (p,n) and (n,p) SCX can similarly be quantified by probability factors.

In addition, rescattering between the knockout and recoil nucleons (i.e. the nucleons of the pair) can also distort the kinematics of the measured events. Previous studies of the deuteron show that, in the kinematics of the current measurement, such internal pair rescattering is strongly suppressed [39].

Thus, the two main reaction mechanisms that effect our measurement are reductions in the measured cross-section due to transparency, and SCX-induced enhancements due to neutron knockout interactions.

One should note that this simple QE picture, with suppressed elastic FSIs, is strongly supported by the fact that it describes well both high-$Q^2$ electron-scattering data and high-energy proton scattering data [15, 45], which have very different reaction mechanisms. In addition, the results of the electron- and proton-scattering experiments give consistent SRC-pair isospin ratios [10, 13, 15] and CM momentum distributions [16, 45, 46].

Non-QE reaction mechanisms such as small-angle leading-nucleon rescattering can modify the measured kinematics. By changing the leading-nucleon momentum, rescattering can cause events with high missing-momentum that originate from interactions with low initial-momentum nucleons. These effects are not accounted for by the SCX and transparency corrections detailed below and can therefore interfere with the interpretation of the data.

Such effects have characteristic kinematics and are expected to increase with nuclear size (i.e. be larger for heavier nuclei). As detailed in the online supplementary materials, we do not observe the expected characteristic behaviors and do observe that the data measured for different nuclei are very similar (see Extended data Fig. 2 and 3). This indicates that A-dependent effects are small and that our data is not significantly distorted by re-scattering effects that go beyond transparency and SCX. See online supplementary materials for details.

**GCF Model: Input Parameters, reaction mechanism corrections, and comparison with experimental data**

The GCF cross-section model and its calculation method are detailed in the online supplementary materials. Here we discuss the model input parameters, corrections for FSI and SCX effects, and assessment of the model systematic uncertainty.

The GCF cross-section calculation requires four external inputs:

1. Nuclear contact values ($C_{NN}^{\alpha}$): For the AV18, AV4', and N2LO we use nuclear contacts that were previously extracted from analyses of two-nucleon momentum distributions [24, 25], obtained from many-body Quantum Monte-Carlo calculations for C [47, 48]. Because we normalize the simulated event yields to the integrated number of $(e,e'p)$ data events, our calculations are only sensitive to the relative values of the contacts.
2. Universal $\tilde{\varphi}_{12}^{\alpha}(p_{12})$ functions: These are taken as the solution of the two-body Schrodinger equation for nucleon pair 1-2 with quantum numbers $\alpha$, see Refs. [23-25] for details. $\tilde{\varphi}_{12}^{\alpha}(\boldsymbol{p}_{12})$ are nucleus-independent, but depend on the *NN* interaction model used in its calculation. In the case of the spin-1 ($s = 1$) quantum state this amounts to the deuteron wave-function shown in Extended Data Fig. 6 (a). For the spin-0 ($s = 0$) quantum state it is the zero-energy solution of the two-body *NN* system, see Extended Data Fig. 6 (b) for the pp channel.
3. SRC pairs center-of-mass momentum distributions: These distributions were studied both theoretically [49, 50] and experimentally [11, 16, 45, 46] and were found to be well described by a three-dimensional gaussian that is defined by its width. For the nuclei considered here, both measurements and theoretical calculations show this width to be about 150 ± 20 MeV/c [16].
4. Excitation energy of the A-2 system: Unlike the other inputs mentioned above, $E_{A-2}^{*}$ was never measured before and can therefore take any value up to about the Fermi-energy (~ 30 MeV).

The comparison of the GCF model with experimental data is done using Monte Carlo integration. We implemented the GCF cross-section model (both the regular and light-cone versions [9, 15, 51-53]) into an event generator that simulates the reaction shown in Extended Data Fig. 1, in which an electron has a hard scattering from a nucleon in an SRC pair within a nucleus, causing both the struck nucleon and the correlated partner nucleon to be ejected from the nucleus. Events are weighted by the GCF cross-section and we account for electron radiation effects using the peaking approximation detailed in Ref. [54], where the radiated photon is emitted in the direction of either the incoming or outgoing electron. See online supplementary materials for details.

To compare our event generator to data, we take the following steps:

- Generate Monte Carlo events as explained above,
- Multiply the weight of each event by the CLAS detection efficiency for the particles detected in that event,
- Smear the generated electron and proton momenta to account for the CLAS resolution,
- Reject events with particles that would not have been detected by CLAS,
- Apply the same event selection cuts used to select data-events.

We accounted for transparency and single-charge exchange (SCX) following Refs. [13, 18] by constructing the following relations:

Eq. 1
$$\sigma^{Exp}_{A(e,e'pp)} = \sigma^{GCF}_{A(e,e'pp)} \cdot P^{pp}_A \cdot T^{NN}_A + \sigma^{GCF}_{A(e,e'np)} \cdot P^{[n]p}_A \cdot T^{NN}_A + \sigma^{GCF}_{A(e,e'pn)} \cdot P^{p[n]}_A \cdot T^{NN}_A,$$

$$\sigma^{Exp}_{A(e,e'p)} = \left(\sigma^{GCF}_{A(e,e'pp)} + \sigma^{GCF}_{A(e,e'pn)}\right) \cdot P^{pp}_A \cdot T^{N}_A + \sigma^{GCF}_{A(e,e'np)} \cdot P^{[n]p}_A \cdot T^{N}_A + \sigma^{GCF}_{A(e,e'nn)} \cdot P^{[n]n}_A \cdot T^{N}_A,$$

where $\sigma^{GCF}_X$ are the GCF simulated events for process X without FSI or SCX, and the $P_A$ and $T_A$ factors are multiplied to the event weights to account for SCX and transparency probabilities, respectively. The $P_A$ and $T_A$ factors do not affect the kinematics of the calculated events.

$T^{NN}_A$ refers to the transparency for both the leading and recoil nucleons being emitted simultaneously, while $T^{N}_A$ refers to the transparency for the leading nucleon independent of the recoil nucleon. We assume that the transparencies for protons and neutrons are the same, and therefore independent of SCX.

As SCX probabilities are different for protons and neutrons and for high and low momentum, the *NN* superscript notation in the P factor marks the exact process being considered, such that particle with (without) square brackets are the ones that undergo (do not undergo) SCX. For example $P^{[p]p}_A$ is the probability that a leading proton in a *pp* pair undergoes SCX, $P^{p[p]}_A$ is this probability for the recoil proton, and $P^{pp}_A = 1 - P^{[p]p}_A - P^{p[p]}_A$ is the probability that no proton undergoes SCX. As can be seen, SCX changes final state neutrons to protons and vice versa. We neglect cases where more than one particle undergoes SCX as these have negligible probability.

The values used for these probabilities are listed in Supplementary Information Table I. They are based on the Glauber calculations of Ref. [18], which agree well with experimental data [41-43]. Both the Glauber calculation and data show that, for the kinematics of the current measurements, these probabilities are energy independent for the leading nucleon. The leading nucleon transparency forces the (*e,e'pp*) reaction to take place near the nuclear surface such that the energy dependence of the recoil nucleon transparency is also expected to be very small. As these effects are model-dependent, we chose to include them in the calculated cross sections, leaving the data fully model independent.

**GCF Model: Systematic Uncertainties**

Uncertainty on the GCF and event-generator input parameters (e.g., CLAS resolution factors, transparency factors, SCX probabilities, Nuclear Contacts, SRC pair center-of-mass motion, A−2 system excitation energy, and the pair relative momentum value for the onset of the SRC regime) all contribute to the total systematic uncertainty of the calculation. We accounted for that by simulating a large number of "universes", in which these input parameters are each randomly drawn from prior probability distributions. We then examined the spread of the generator results across this space of universes to produce a systematic uncertainty band that captures 68% of the examined parameter combinations.

The following values and Gaussian uncertainties were used for these parameters for $^{12}$C:

- $\sigma_{CM}$, the gaussian width of the SRC pair center-of-mass momentum distribution [16]: 150 ± 20 MeV/c.
- The nuclear contacts for AV18, AV4' and N2LO are taken from Ref. [25], specifically the *k*-space fits in their supplementary materials Table 1.
- SCX and nuclear transparency probabilities and uncertainties (Supplementary Information Table I) are taken from Ref. [13]
- $E^*_{A-2}$, the excitation energy of the residual A − 2 system, was varied uniformly between 0 − 30 MeV.
- The $p^{min}_{rel}$ cutoff in the universal two-body functions was varied uniformly between 250 and 350 MeV/c.
- The simulated electron resolution was varied uniformly between 1.0 − 1.5%.
- The simulated proton resolution was varied uniformly between 0.8 − 1.2%.

- The off-shell electron-nucleon cross section was chosen to be either $\sigma_{CC1}$ or $\sigma_{CC2}$ from Ref. [21].

We note that while the individual parameter uncertainties are independent of each other, their impact on the calculated cross-section can be correlated. e.g. both $E^*_{A-2}$ and $\sigma_{CM}$ can affect the initial pair energy in a similar manner (Extended Data Fig. 1). By varying all parameters simultaneously we include the effects of such correlations. We also note that the use of $\sigma_{CC1}$ vs. $\sigma_{CC2}$ is a discrete choice, in contrast to all other variations that are continuous. We compared the calculations done using only $\sigma_{CC1}$ and only $\sigma_{CC2}$ and do not see a significant difference in the resulting distributions.

**Initial Nucleon Energy:**

We use the convention that the spectral function depends on $\epsilon_i$, the initial off-shell energy of the struck nucleon prior to scattering. The expected initial off-shell energy for nucleons in a stationary pair is given by:

$$\text{Eq. 2} \qquad \epsilon_i = m_A - m_{A-2} - \sqrt{p_i^2 + m_N^2},$$

which is shown by red arrows in Fig. 3 and Extended Data Fig. 7 (d - k).

**Scalar Limit Estimation:**

The general expectation for a fully scalar *NN* interaction and a symmetric nucleus, is that the abundance of pairs will be equal for all isospin, spin, and spin-projection states. This implies that the number of spin-1 *pn*-SRC pairs should be three times the number of spin-0 *pp*-, *pn*-, and *nn*-pairs due to the three possible spin orientations. Therefore, simple counting implies:

$$\text{Eq. 3} \qquad \frac{\#e'pp}{\#e'p} = \frac{2N_{pp}^{s=0}}{2N_{pp}^{s=0}+N_{pn}^{s=0}+N_{pn}^{s=1}} = \frac{2N^{s=0}}{(2+1+3)N^{s=0}} = \frac{1}{3},$$

where $N_{NN}^{s=S}$ is the number of *NN* pairs in a spin S state. This limit is shown as the dashed line labeled ‘scalar limit’ in Extended Data Fig. 6 (c).

**Cutoff Dependence and Non-Local Chiral Interactions:**

In addition to the local interactions studied in this work, nuclear structure calculations are often performed using non-local interactions, which feature different high-momentum asymptotic behavior as compared to local ones. The non-local versions of the $\chi$EFT interactions have momentum-space cutoffs and are considered to be ‘softer’ than the local interactions studied here.

The main limitation for studying such interactions using the GCF framework presented here is that, at the moment, there are no available calculations of the two-nucleon momentum distribution in $^{12}C$ using these interactions. Therefore, we are unable to determine the nuclear contacts for these interactions in a fully theoretical fashion as is done for the local interactions considered above.

One previous work [22] studied the non-local N3LO(600 MeV/c) interaction [5] using the GCF by extracting the ratio of spin-1 to spin-0 contacts from a fit to the experimental data of Refs. [10, 11]. While this procedure cannot be compared on an equal footing with the fully theoretical predictions we have for the local interactions, it is still interesting to see how they compare with each other and with the data. This comparison is shown in Extended Data Fig. 8, which is equivalent to Fig. 2 (a) and 3 (a,b) of the main text. As can be seen, the non-local N3LO(600 MeV/*c*) interaction seems to reproduce the experimental data well up to its cutoff, but then decays faster than the local interactions. This is an encouraging observation as the 600 MeV/*c* cutoff of this interaction is well above the 300 MeV/*c* c.m. momentum cutoff of the *NN* phase-shift data used in its construction.

The predictions of the N3LO(600 MeV/c) interaction are quite similar to those of the N2LO(1.2 fm) interaction. This might seem surprising as the 1.2 fm cutoff corresponds to momenta of ~400 MeV/*c* [26, 56] that is smaller than 600 MeV/c and, as mentioned in the main text, one expects large errors in the predictions of the different $\chi$EFT interactions already at their cutoff scale. Our observations, consistent with theoretical expectations, indicate that for the processes being studied here the nature of the position-space regulators makes their effects significant only at a relatively high momenta scale of 600 – 700 MeV/*c*, see Fig. 3 (d) – (i).

Future studies will focus on using the experimental data provided in this work (which is much more detailed than that of Refs. [10, 11]) to fit the nuclear contacts for different local and non-local interactions and study the dependence of the results on the chiral expansion order and cutoff.

For completeness, we note that from a theoretical standpoint, the reaction diagram used for the GCF calculations and shown in Extended Data Fig. 1 can be viewed as a ‘high-resolution’ starting point for a unitary-transformed calculation [55]. As a thought exercise, the $\chi$EFT NN interactions used here can be considered as resulting from applying unitary transformations to models that have shorter distance / higher-momentum cutoffs. As this process would introduce many-body interaction currents to the description of the electron scattering reaction, the use of a high-resolution (one-body) reaction description with $\chi$EFT interactions, as done in this work, is non-trivial. This is one explanation for the disagreement between the data and calculations at high $p_{miss}$. The data presented here can therefore quantify the importance of such many-body effects and demonstrate that they become significant only above the cutoff for non-relativistic calculations and at much higher momenta when relativistic effects are accounted for. This can help guide future studies of effects such as relativity and non-nucleonic degrees of freedom.

**Fitting Contact Ratios to (e,e'pp)/(e,e'p) data:**

In addition to using the $^{12}$C contact values determined by Ref. [25] to make GCF predictions, we also used the measured *(e,e'pp)/(e,e'p)* data (Fig. 2b) to infer the ratio of spin-0 to spin-1 contacts, $C_{pp}^{s=0}/C_{pn}^{s=1}$, for each measured nucleus. We specifically used *(e,e'pp)/(e,e'p)* ratio as cut-off effects largely cancel. We employed Bayesian inference with the contact ratio being the sole parameter of interest. All other GCF parameters were treated as nuisance parameters and were integrated out, with prior distributions matching the systematic uncertainty sampling distributions described above. We assumed a scale-invariant prior on the ratio of interest, $C_{pp}^{s=0}/C_{pn}^{s=1}$. Additionally, we assume a ratio of spin-0 contacts, $C_{pp}^{s=0}/C_{pn}^{s=0}$ of 1 for C, Al, and Fe. For Pb, this ratio was assumed to equal 82 / 126 = 0.65.

To evaluate the likelihood of the data for a given set of values of the GCF parameters, we multiplied the individual likelihoods of the 10 data points in Fig. 2b, assuming the data were distributed normally. The posterior distribution was determined by scanning over $C_{pp}^{s=0}/C_{pn}^{s=1}$ and Monte Carlo integrating over all nuisance parameters at each step.

The results are presented in Extended Data Table 1. The central value is the maximum (mode) of the posterior distribution while the uncertainty intervals are the smallest intervals containing 68.3% (1σ) and 95.5% (2σ) of the total posterior. The table also shows the double ratio of contacts for nucleus A relative to C which is independent of the *NN* interaction model [25].

**Additional References:**

[31] "A Double Target System for Precision Measurements of Nuclear Medium Effects", H. Hakobyan et al., Nucl. Inst. and Meth. A **592**, 218-223 (2008).

[32] "Short range correlations and the isospin dependence of nuclear correlation functions", R. Cruz-Torres, A. Schmidt, G.A. Miller, L.B. Weinstein, N. Barnea, R. Weiss, E. Piasetzky, and O. Hen, Phys. Lett. B 785, 304-308 (2018).

[33] "Short-range correlations and the charge density", R. Weiss, A. Schmidt. G.A. Miller, and N. Barnea, Phys. Lett. B **790**, 484-489 (2019).

[34] "Exclusive electrodisintegration of 3He at high *Q*2 II. Decay function formalism" M.M. Sargsian, T. V. Abrahamyan, M. I. Strikman, and L. L. Frankfurt, Phys. Rev. C 71, 044615 (2005).

[35] "Feynman graphs and generalized eikonal approach to high energy knock-out processes", L.L. Frankfurt, M.M. Sargsian, and M.I. Strikman, Phys. Rev. C **56**, 1124 (1997).

[36] "Quasielastic 3He(e,e'p)2H Reaction at *Q*2 = 1.5 GeV2 for Recoil Momenta up to 1 GeV/c", M. Rvachev et al. Phys. Rev. Lett. **94**, 192302 (2005).

[37] "Measurement of the 3He(e,e'p)pn Reaction at High Missing Energies and Momenta", F. Benmokhtar et al. Phys. Rev. Lett. **94**, 082305 (2005).

[38] "Experimental Study of Exclusive $^{2}$H(e,e'p)n Reaction Mechanisms", K. S. Egiyan et al. (CLAS Collaboration), Phys. Rev. Lett. **98**, 262502 (2007).

[39] "Probing the High Momentum Component of the Deuteron at High $Q^2$", W.U. Boeglin et al. Phys. Rev. Lett. **107**, 262501 (2011).

[40] "Color transparency: past, present and future", D. Dutta, K. Hafidi, and M. Strikman, Prog. Part. Nucl. Phys. **69**, 1-27 (2013).

[41] "Measurement of Transparency Ratios for Protons from Short-Range Correlated Pairs", O. Hen et al. (CLAS Collaboration), Phys. Lett. B **722**, 63-68 (2013).

[42] "Measurement of Nuclear Transparency Ratios for Protons and Neutrons", M. Duer, et al. (CLAS Collaboration), Phys. Lett. B **797**, 134792 (2019).

[43] "Extracting the mass dependence and quantum numbers of short-range correlated pairs from A(e,e'p) and A(e,e'pp) scattering", C. Colle et al., Phys. Rev. C **92**, 024604 (2015).

[44] "Universality of many-body two-nucleon momentum distributions: Correlated nucleon spectral function of complex nuclei", C. Ciofi degli Atti and H. Morita, Phys. Rev. C **96**, 064317 (2017).

[45] "n-p short range correlations from (p,2p + n) measurements", A. Tang et al (EVA Collaboration), Phys. Rev. Lett. **90**, 042301 (2003).

[46] “Investigation of proton-proton short-range correlations via the 12C(e,e’pp) reaction”, R. Shneor et al. (Jefferson Lab Hall A Collaboration), Phys. Rev. Lett. **99**, 072501 (2007).
[47] “Single- and two-nucleon momentum distributions for local chiral interactions”, D. Lonardoni, S. Gandolfi, X. B. Wang, and J. Carlson, Phys. Rev. C **98**, 014322 (2018).
[48] “Nucleon and nucleon-pair momentum distributions in A ≤ 12 nuclei”, R.B. Wiringa, R. Schiavilla, S.C. Pieper, and J. Carlson, Phys. Rev. C **89**, 024305 (2014).
[49] “Realistic model of the nucleon spectral function in few and many nucleon systems”, C. Ciofi degli Atti and S. Simula, Phys. Rev. C **53**, 1689 (1996).
[50] “Factorization of exclusive electron-induced two-nucleon knockout”, C. Colle, W. Cosyn, J. Ryckebusch, and M. Vanhalst, Phys. Rev. C **89**, 024603 (2014).
[51] “Short range correlations in nuclei as seen in hard nuclear reactions and light cone dynamics”, L. Frankfurt and M. Strikman, Modern topics in electron scattering, B. Frois (ed.), I. Sick (ed.), 645-694, (1992).
[52] “Multinucleon short-range correlation model for nuclear spectral functions: Theoretical framework”, O. Artiles and M. Sargsian, Phys. Rev. C **94**, 064318 (2016).
[53] “Relation between equal-time and light-front wave functions”, G.A. Miller and B.C. Tiburzi, Phys. Rev. C **81**, 035201 (2010).
[54] “Radiative corrections for (e,e’p) reactions at GeV energies”, R. Ent, B.W. Filippone, N.C.R. Makins, R.G. Milner, T.G. O’Neill, and D.A. Wasson, Phys. Rev. C **64**, 054610 (2001).
[55] “Scale dependence of deuteron electrodisintegration”, S. N. More, S. K. Bogner, and R. J. Furnstahl, Phys. Rev. C **96**, 054004 (2017).
[56] “Quantum Monte Carlo calculations of light nuclei with local chiral two- and three-nucleon interactions”, J. E. Lynn, I. Tews, J. Carlson, S. Gandolfi, A. Gezerlis, K. E. Schmidt, and A. Schwenk, Phys. Rev. C **96**, 054007 (2017)

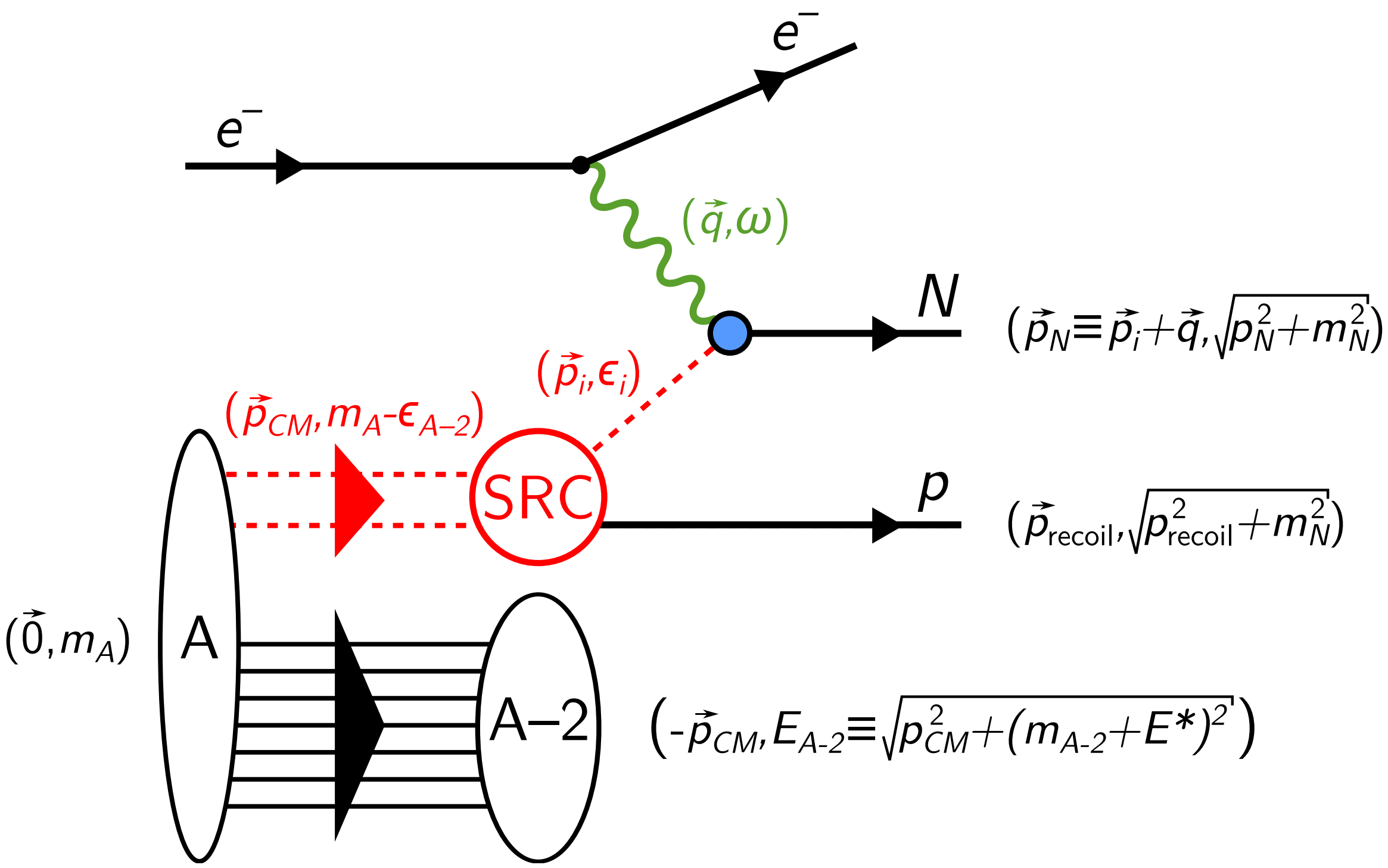

**Extended Data Fig. 1 | SRC pair breakup:** Diagrammatic representation and 4-momentum kinematics of the two-nucleon knockout $A(e,e'Np)$ reaction, within the SRC model. Dashed red lines represent off-shell particles and solid black lines represent detected particles. The A-2 system is undetected.

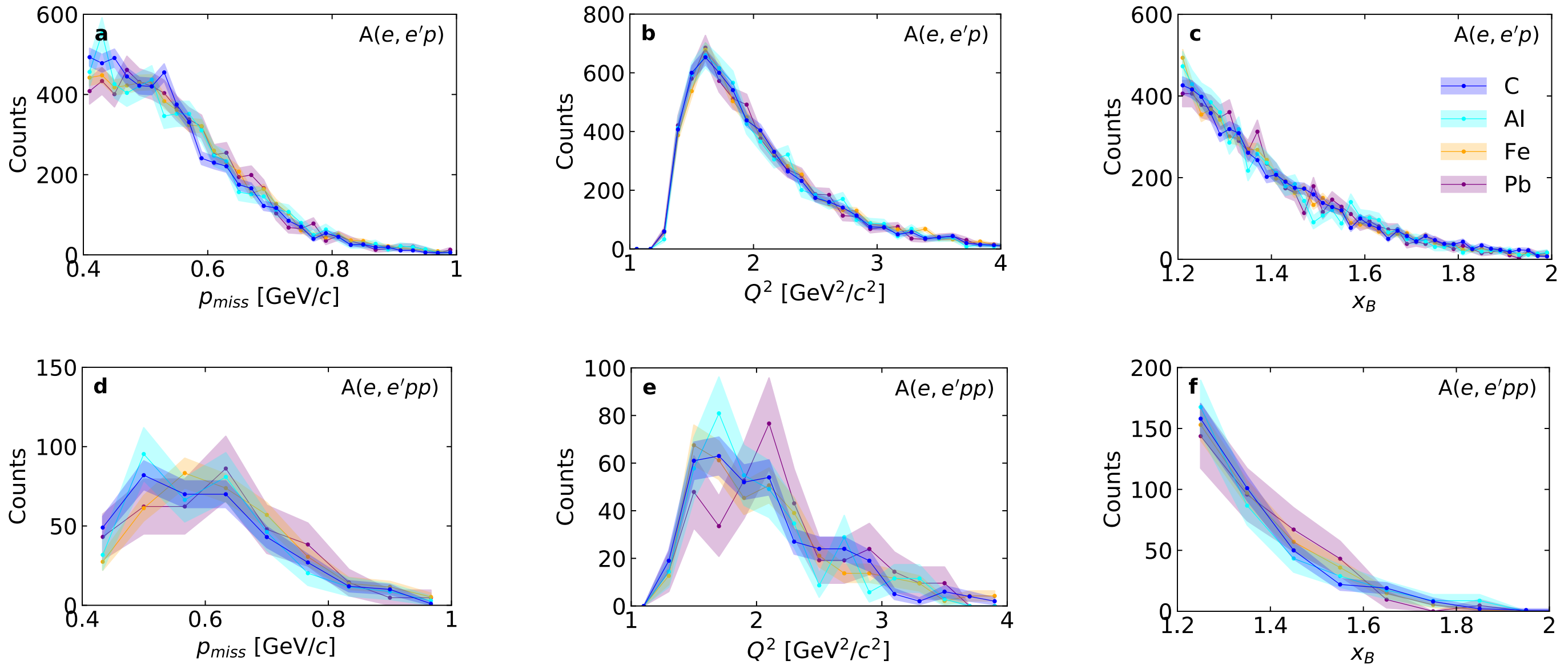

**Extended Data Fig. 2 | Kinematical distributions and A(e,e'pp) / A(e,e'p) ratios for A = 12 – 208 Nuclei:** Comparison of the number of events plotted versus $(e,e'p)$ missing momentum (c, f), $Q^2$ (d, g) and $x_B$ (e, h) for A$(e,e'p)$ (c-e) and A$(e,e'pp)$ (f-h) reactions. The total number of counts in Al (light blue), Fe (orange), and Pb (purple) was scaled to match that of C (dark blue). The bands indicate the statistical uncertainty of the data.

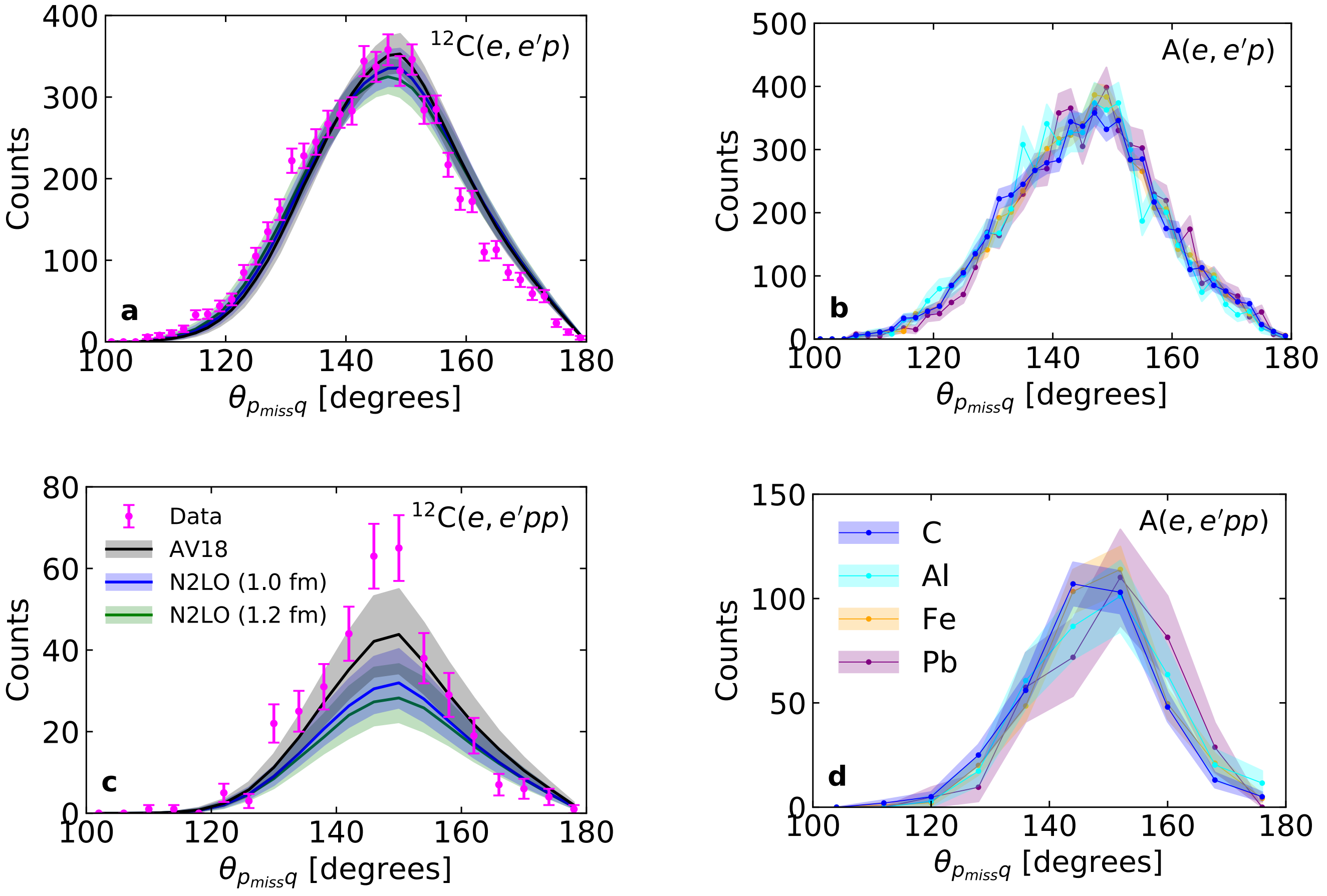

**Extended Data Fig. 3 | Momentum-transfer and missing-momentum angular correlations:** Distribution of the relative angle between the momentum transfer ***q*** and the (*e,e'p*) missing momentum for A(*e,e'p*) (top) and A(*e,e'pp*) (bottom) reactions. Left: Comparison of $^{12}$C data and GCF calculations using different *NN* interaction models (colored bands). Right: Comparison of data for C (blue), Al (light blue), Fe (orange), and Pb (purple) nuclei. The width of the band and the data error bars show the model systematic uncertainties and data statistical uncertainties, respectively, each at the 1σ or 68% confidence level. The total number of counts in the Al, Fe and Pb data has been scaled to match C. The bands in panels b and d indicate the statistical uncertainty of the data.

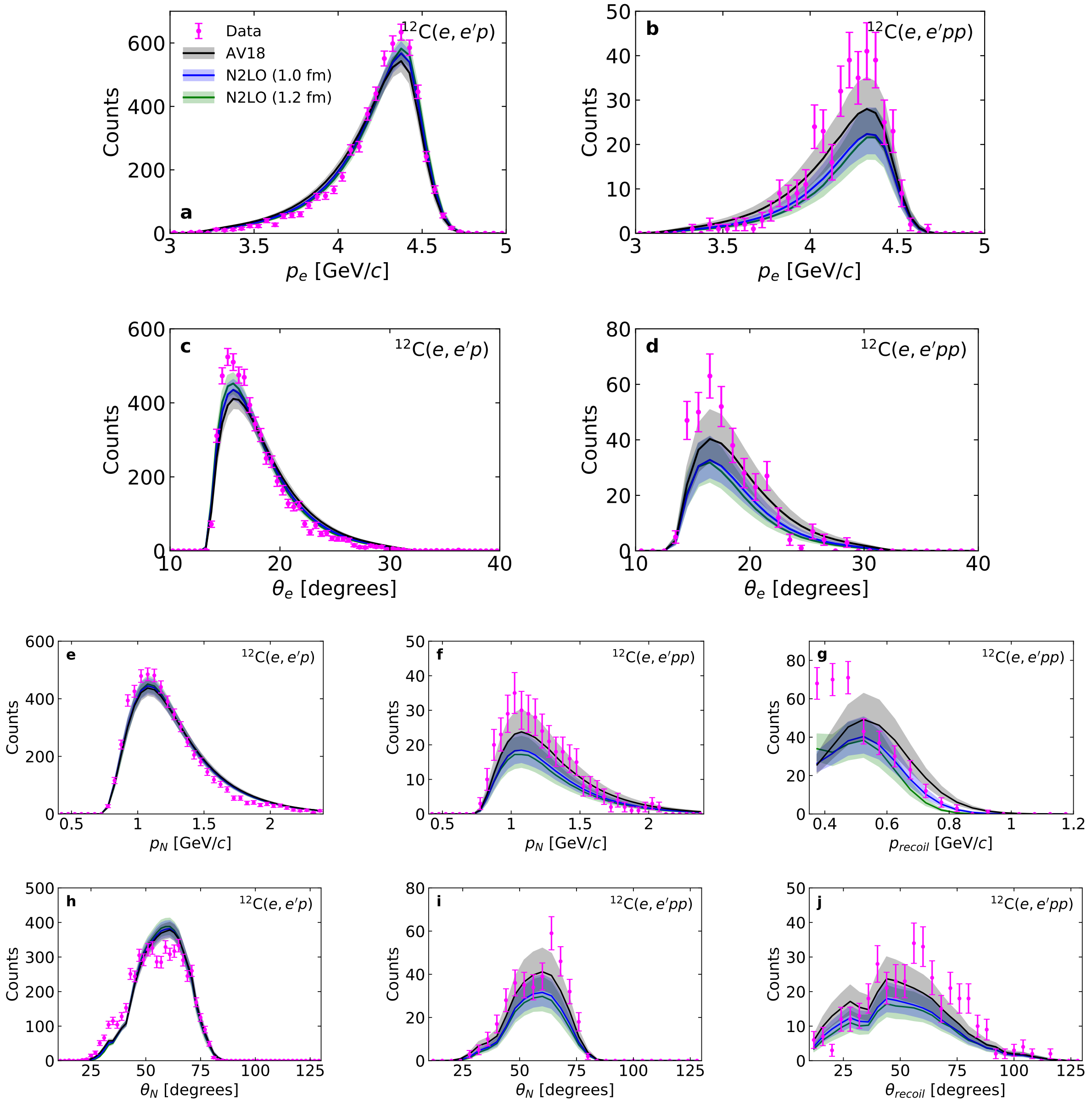

**Extended Data Fig. 4 | Electron and proton kinematics:** Scattered-electron and proton momentum and angle distributions for $^{12}$C($e,e'p$) (a, c, e, h) and $^{12}$C($e,e'pp$) (b, d, f, g, i, j) events. Colored bands show GCF calculations. The width of the band and the data error bars show the model systematic uncertainties and data statistical uncertainties, respectively, each at the 1σ or 68% confidence level.

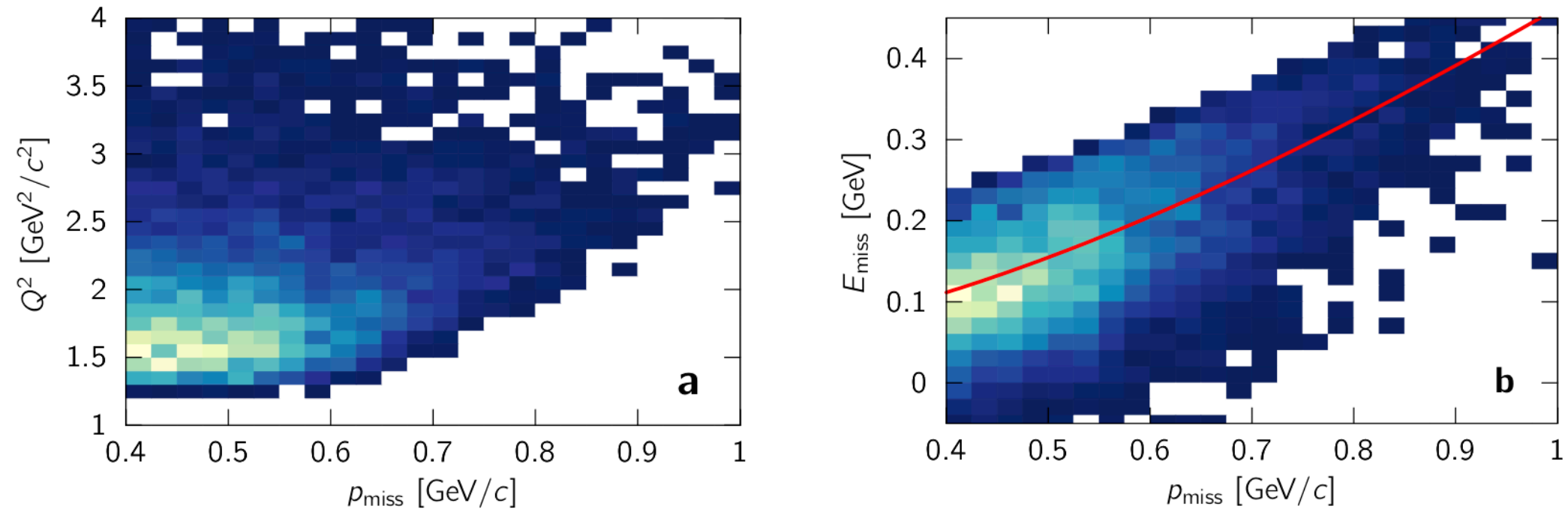

**Extended Data Fig. 5 | Kinematic Correlations of $^{12}$C(e,e'p) Events.** Q2 versus missing momentum distribution of $^{12}$C(e,e'p) data (a). Due to the event selection criteria, as $P_{miss}$ approaches 1 GeV/c, the minimum $Q^2$ of the data approaches 3 GeV/c . $E_{miss}$ versus $P_{miss}$ of $^{12}$C(e,e'p) data (b). The red line indicates the expected $E_{miss}$-$P_{miss}$ correlation for the break-up of a stationary pair.

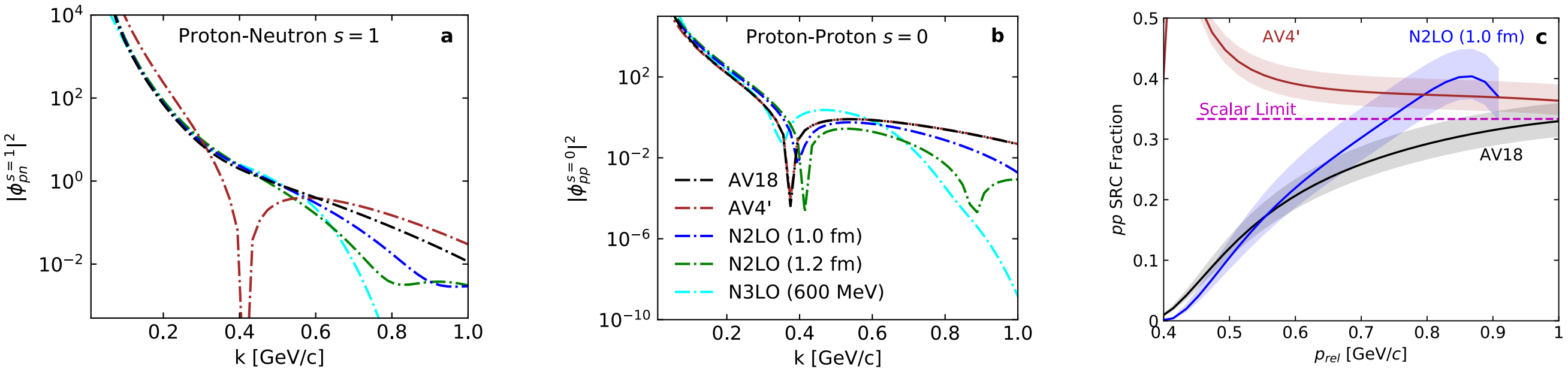

**Extended Data Fig. 6 | Universal functions for *pp* and *np* pairs and the momentum dependence of their ratio.** Proton-neutron (a) and proton-proton (b) relative momentum distributions for different *NN* interaction models studied in this work as well as the momentum dependence of the fraction of protons belonging to *pp*-SRC pairs in $^{12}$C (c).

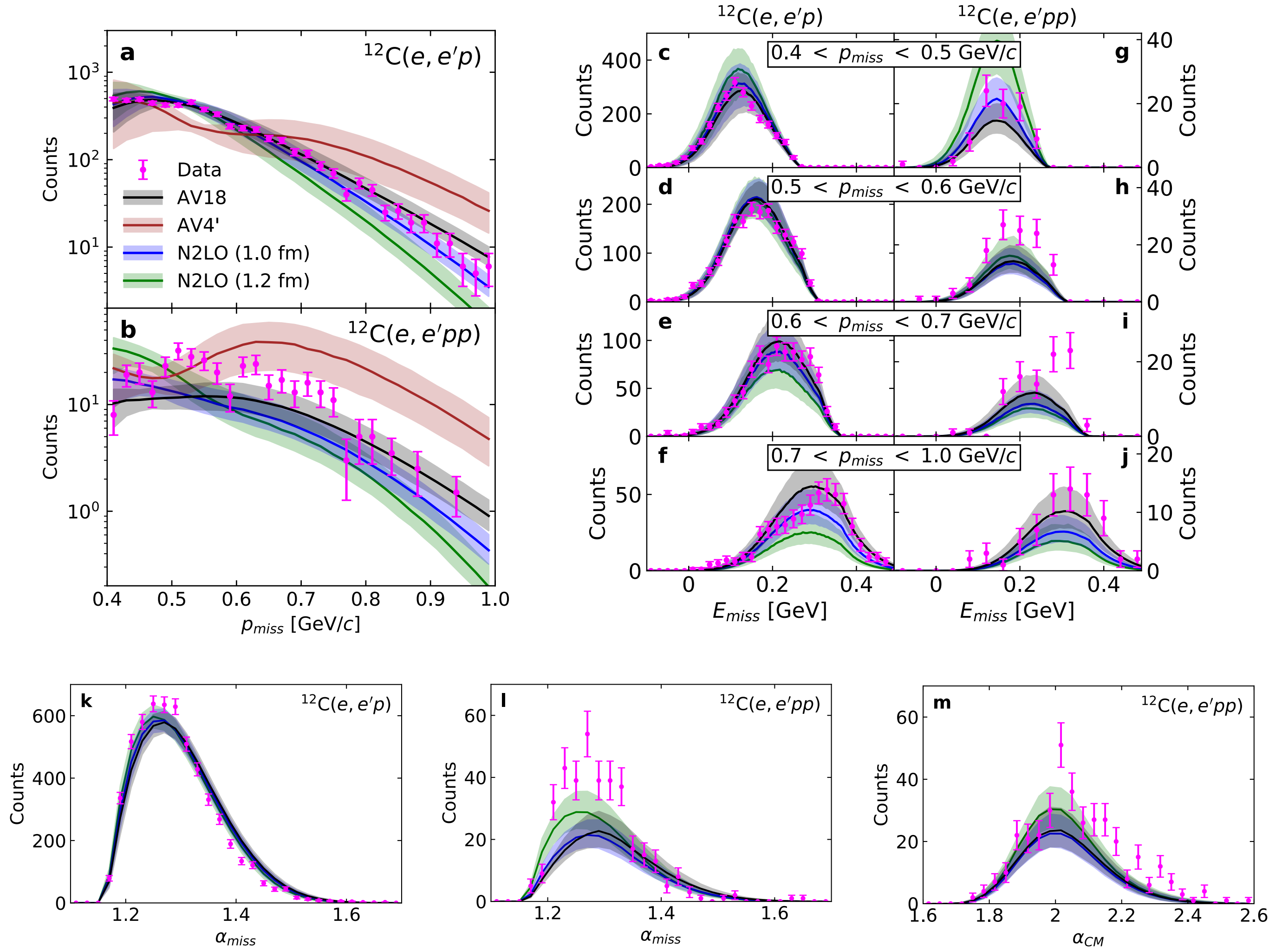

**Extended Data Fig. 7 | LC Calculations of the nuclear spectral function and momentum fractions.** Same as Fig. 3 (a) – (j), with the addition of reconstructed initial light-cone momentum fraction carried by the struck nucleon for (*e,e'p*) (k) and (*e,e'pp*) (l) events as well as the total pair light-cone momentum fraction for (*e,e'pp*) events (m). Bands show the results of GCF calculations using LC formalism and various NN interaction models. The width of the band and the data error bars show the model systematic uncertainties and data statistical uncertainties, respectively, each at the 1σ or 68% confidence level.

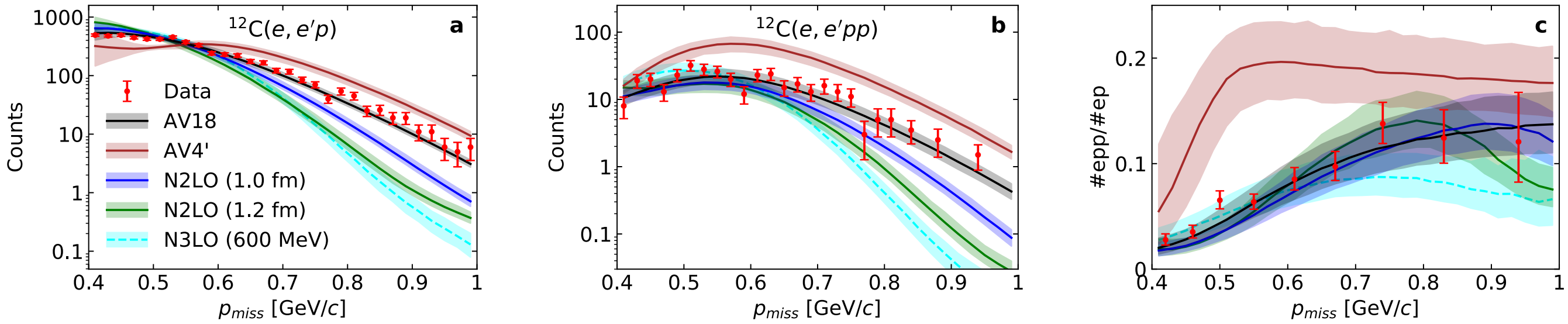

**Extended Data Fig. 8 | Scale dependence and non-local Interactions:** Same as Fig. 2 (a) and 3 (a,b), but including the non-local N3LO(600) interaction also. The width of the band and the data error bars show the model systematic uncertainties and data statistical uncertainties, respectively, each at the 1σ or 68% confidence level. See Methods for details.

| Interaction | $^{12}$C | | $^{27}$Al [%] | Al/C | $^{56}$Fe [%] | Fe/C | $^{208}$Pb [%] | Pb/C |
|---|---|---|---|---|---|---|---|---|
| | This Work [%] | Ref. [25] | | | | | | |
| AV18 | $6.7^{+1.7\,(4.7)}_{-1.0\,(1.7)}$ | 7.2±1.6 | $4.5^{+1.9\,(7.0)}_{-1.0\,(1.5)}$ | 0.67 | $6.4^{+1.7\,(4.8)}_{-1.0\,(1.8)}$ | 0.96 | $1.4^{+0.9\,(3.0)}_{-0.5\,(0.7)}$ | 0.21 |
| N2LO (1.0 fm) | $8.1^{+2.8\,(9.7)}_{-1.4\,(2.3)}$ | 8.2±1.2 | $5.3^{+2.7\,(11.3)}_{-1.2\,(2.0)}$ | 0.65 | $8.0^{+2.8\,(9.2)}_{-1.5\,(2.4)}$ | 0.99 | $1.8^{+1.2\,(4.4)}_{-0.6\,(0.8)}$ | 0.22 |
| N2LO (1.2 fm) | $15.1^{+4.4\,(13.6)}_{-2.5\,(4.1)}$ | 18.7±2.8 | $9.7^{+4.3\,(14.2)}_{-2.2\,(3.5)}$ | 0.64 | $13.2^{+5.0\,(11.5)}_{-2.5\,(4.0)}$ | 0.87 | $3.6^{+2.1\,(6.4)}_{-1.2\,(1.8)}$ | 0.24 |

**Extended Data Table 1 | Extracted Contact Ratios:** $C_{pp}^{s=0}/C_{np}^{s=1}$ for different nuclei as extracted by fitting the GCF calculation to the *A(e,e'pp)/A(e,e'p)* data shown in Fig. 2b. The uncertainties are shown at the 1σ (2σ) or 68% (95.5%) confidence level.

# Supplementary Materials

## The GCF Model

The derivation of the GCF and its application for describing SRCs in nuclei are detailed in Refs. [22-25, 32, 33]. It is applicable at missing momenta where SRC dominates [24, 25]. Here we review its main characteristics that are relevant for this work.

The GCF is a generalization of the atomic contact formalism, successfully used to describe scale-separated, strongly interacting, two-component Fermi systems. It assumes that at very high momenta, the asymptotic nuclear many-body ground state wave function $\Psi^A$ can be factorized into an SRC pair and a residual A-2 system as [24, 25]:

$$\text{Eq. 1} \qquad \Psi^A \xrightarrow{p_{12}\to\infty} \sum_\alpha \tilde{\varphi}_{12}^\alpha(\boldsymbol{p}_{12})\tilde{A}_{12}^\alpha\left(\boldsymbol{P}_{12}, \{\boldsymbol{p}_k\}_{k\neq 1,2}\right),$$

where $\alpha$ denotes the SRC pair quantum numbers, $\tilde{\varphi}_{12}^\alpha(p_{12})$ are universal two-body functions of the relative momentum of the SRC pair $\boldsymbol{p}_{12} = (\boldsymbol{p}_1 - \boldsymbol{p}_2)/2$, and $\tilde{A}_{12}^\alpha$ describes the motion of the A-2 system with total pair CM motion, $\boldsymbol{P}_{12} = (\boldsymbol{p}_1 + \boldsymbol{p}_2)$. $\tilde{\varphi}_{12}^\alpha(p_{12})$ are normalized such that the integral over $|\tilde{\varphi}_{12}^\alpha(p_{12})|^2$ from $k_F$ to infinity equals 1 [24, 25].

Under this approximation, the asymptotic high-momentum proton spectral function can be written as a sum over SRC pairs [22]:

$$\text{Eq. 2} \qquad S^p(p_i, \epsilon_i) = C_{pn}^{s=1} S_{pn}^{s=1}(p_i, \epsilon_i) + C_{pn}^{s=0} S_{pn}^{s=0}(p_i, \epsilon_i) + 2C_{pp}^{s=0} S_{pp}^{s=0}(p_i, \epsilon_i),$$

where $C_{NN}^\alpha$ are the nuclear contacts, which measure the probability to find a proton-proton (pp) pair or a proton-neutron (pn) pair with quantum numbers α close together. The functions $S_{NN}^\alpha(p_i, \epsilon_i)$ are the individual contributions of these pairs to the total spectral function. The $s = 1$ state corresponds to the spin-one deuteron state, and $s = 0$ corresponds to the spin-zero s-wave state.

The single-pair spectral function $S_{NN}^\alpha(p_i, \epsilon_i)$ is given by a convolution of its relative and CM motion:

$$\text{Eq. 3} \qquad S_{NN}^\alpha(p_i, \epsilon_i) = \frac{1}{4\pi}\int \frac{d\boldsymbol{p}_{recoil}}{(2\pi)^3} \delta\left(f(\boldsymbol{p}_{recoil})\right) \Theta\left(\boldsymbol{p}_{rel} - p_{rel}^{min}\right) |\tilde{\varphi}_{NN}^\alpha(\boldsymbol{p}_{rel})|^2 n_{NN}^\alpha(\boldsymbol{P}_{CM}),$$

and

$$\text{Eq. 4} \qquad f(\boldsymbol{p}_{recoil}) = \epsilon_i + \epsilon_{\text{recoil}} - m_A + \sqrt{P_{CM}^2 + (m_{A-2} + E_{A-2}^*)^2} \qquad ,$$

where $\boldsymbol{P}_{CM} \equiv \boldsymbol{p}_i + \boldsymbol{p}_{recoil}, \boldsymbol{p}_{rel} \equiv (\boldsymbol{p}_i - \boldsymbol{p}_{recoil})/2$, $n_{NN}^\alpha(P_{CM})$ is the CM momentum distribution of the SRC pair given by $C_{NN}^\alpha n_{NN}^\alpha(\boldsymbol{P}_{CM}) = \langle \tilde{A}_{12}^\alpha(\boldsymbol{P}_{CM}) | \tilde{A}_{12}^\alpha(\boldsymbol{P}_{CM})\rangle$, $m_A$ and $m_{A-2}$ are the ground state masses of the initial A and final A-2 nuclear systems, $E_{A-2}^*$ is the excitation energy of the A-2 system, and $\epsilon_{\text{recoil}}$ is the energy of the recoil nucleon. We assume that the pair CM momentum distribution is the same for all pairs: $n_{NN}^\alpha(\boldsymbol{P}_{CM}) = n_{CM}(\boldsymbol{P}_{CM})$. The step function $\Theta\left(\boldsymbol{p}_{rel} - p_{rel}^{min}\right)$ ensured that we only integrate over pairs with large relative momentum, since the GCF models only SRC pairs that are expected to dominate above $k_F$. We note that while the universal functions used here are non-relativistic, we use relativistic expressions for the nucleon energies since the kinematics of the measured high-$Q^2$ processes is highly relativistic.

The integrand of Eq. 3 is called the two-body decay function $D_A(\boldsymbol{p}_i, \boldsymbol{p}_{recoil}, E_R)$, which represents the probability for a hard knockout of a nucleon with initial momentum $\boldsymbol{p}_i$, followed by an emission of a recoil nucleon with momentum $\boldsymbol{p}_{recoil}$ [9, 15, 17, 34]. $E_R$ is the energy of the A−1 system, composed of the recoil nucleon and the residual A − 2 nucleus. Integrating the decay function over all recoil nucleon momenta ($\boldsymbol{p}_{recoil}$) yields the spectral function.

The GCF model, as presented above, requires four external inputs:

5. <u>Nuclear contact values ($C_{NN}^\alpha$)</u>: For the AV18, AV4', and N2LO we use nuclear contacts that were previously extracted from analyses of two-nucleon momentum distributions [24, 25], obtained from many-body Quantum Monte-Carlo calculations for C [47, 48]. Because we normalize the simulated event yields to the integrated number of (e,e'p) data events, our calculations are only sensitive to the relative values of the contacts.
6. <u>Universal $\tilde{\varphi}_{12}^\alpha(p_{12})$ functions</u>: These are taken as the solution of the two-body Schrodinger equation for nucleon pair 1-2 with quantum numbers $\alpha$, see Refs. [24, 25] for details. $\tilde{\varphi}_{12}^\alpha(\boldsymbol{p}_{12})$ are nucleus-independent, but depend on the *NN* interaction model used in its calculation. In the case of the spin-1 ($s = 1$) quantum state this amounts to the deuteron wave-function shown in Extended Data Fig. 6 (a). For the spin-0 ($s = 0$) quantum state it is the zero-energy solution of the two-body *NN* system, see Extended Data Fig. 6 (b) for the pp channel.

7. SRC pairs center-of-mass momentum distributions: These distributions were studied both theoretically [49, 50] and experimentally [11, 16, 45, 46] and were found to be well described by a three-dimensional gaussian that is defined by its width. For the nuclei considered here, both measurements and theoretical calculations show this width to be about 150 ± 20 MeV/c [16].
8. Excitation energy of the A-2 system: Unlike the other inputs mentioned above, $E^*_{A-2}$ was never measured before and can therefore take any value up to an order of the Fermi-energy (~ 30 MeV).

We note that as mentioned in main text, calculations of the nuclear spectral function are not feasible for generic nuclear systems. However, for the specific case of three-nucleon system and nuclear matter such calculations are feasible, and their results agree with the model presented above [27]. Additionally, the GCF only includes the effects of two-nucleon correlations and described by the two-body universal functions. Explicit three-nucleon forces and sequential NN interactions can both induce three nucleon correlations that are not described by the GCF as presented above. Previous studies of realistic many-body wave-functions (including three-nucleon forces) show that the GCF can reproduce the calculated momentum distribution in the range relevant for the current study to an accuracy of approximately 10% [24].

**GCF Event Generator, FSI, and SCX Corrections:**

The GCF-based event generator used here simulates the reaction shown in Extended Data Fig. 1, in which an electron has a hard scattering from a nucleon in an SRC pair within a nucleus, causing both the struck nucleon and the correlated partner nucleon to be ejected from the nucleus. The generator generates particles over the full phase-space by sampling events randomly from the probability distribution:

Eq. 5 $$P(Q^2, \mathrm{x_B}, \phi_\mathrm{e}, \boldsymbol{P}_{CM}, \Omega_\mathrm{recoil}) = \frac{1}{\Delta Q^2} \times \frac{1}{\Delta x_B} \times \frac{1}{2\pi} \times n_{CM}(\boldsymbol{P}_{CM}) \times \frac{1}{4\pi},$$

and produces a list of events, each containing momentum vectors for a scattered electron ($\boldsymbol{p}_e$), a leading nucleon ($\boldsymbol{p}_N$), and a recoil nucleon ($\boldsymbol{p}_{recoil}$). The cross section for each event is calculated based on the integrand of Eq. 3 above and is given by:

Eq. 6 $$\frac{\mathrm{d}^8\sigma}{\mathrm{d}Q^2\, \mathrm{dx_B}\, \mathrm{d}\phi_\mathrm{e}\, \mathrm{d}^3\boldsymbol{P}_{CM}\, \mathrm{d}\Omega_\mathrm{recoil}} = \frac{\sigma_{eN}}{32\pi^4} \frac{\epsilon_N \epsilon_{recoil} \boldsymbol{p}^2_{recoil}\ \Theta\left(\boldsymbol{p}_{rel} - p^{min}_{rel}\right) n_{CM}(\boldsymbol{P}_{CM})}{\left|\epsilon_{recoil}\left(\boldsymbol{p}_{recoil} - \boldsymbol{Z} cos\theta_{z,recoil}\right) + \epsilon_N \boldsymbol{p}_{recoil}\right|} \frac{\omega}{2E_{beam}E_e x_B} \sum_\alpha C_\alpha |\tilde{\varphi}^\alpha_{NN}(\boldsymbol{p}_{rel})|^2 \quad ,$$

where $\boldsymbol{Z} \equiv \boldsymbol{q} + \boldsymbol{P}_\mathrm{CM}$ and $\theta_{Z,recoil}$ is the angle between $\boldsymbol{Z}$ and $\boldsymbol{p}_{recoil}$. For the case of pp pairs channels, $C_\alpha$ is equal to twice the nuclear contact for that channel. The starting weight w for each event is then given by $d\sigma/P$:

Eq. 7
$$w = \frac{\sigma_{eN}}{4\pi^2} \Delta Q^2 \Delta x_B \frac{\epsilon_N \epsilon_{recoil} \boldsymbol{p}^2_{recoil}\ \Theta\left(\boldsymbol{p}_{rel} - p^{min}_{rel}\right)}{\left|\epsilon_{recoil}\left(\boldsymbol{p}_{recoil} - \boldsymbol{Z} cos\theta_{z,recoil}\right) + \epsilon_N \boldsymbol{p}_{recoil}\right|} \frac{\omega}{2E_{beam}E_e x_B} \sum_\alpha C_\alpha |\tilde{\varphi}^\alpha_{NN}(\boldsymbol{p}_{rel})|^2.$$

The generation of events also includes electron radiation effects, introduced using the peaking approximation detailed in Ref. [54].

To compare our event generator to data, we take the following steps:

- Generate Monte Carlo events as explained above,
- Multiply the weight of each event by the CLAS detection efficiency for the particles detected in that event,
- Smear the generated electron and proton momenta to account for the CLAS resolution,
- Reject events with particles outside of the fiducial region of detected particles in CLAS,
- Apply the same event selection cuts used to select data-events.

We accounted for transparency and single-charge exchange (SCX) following Refs. [13, 18] by constructing the following relations:

Eq. 8 $$\sigma^{Exp}_{A(e,e'pp)} = \sigma^{GCF}_{A(e,e'pp)} \cdot P^{pp}_A \cdot T^{NN}_A + \sigma^{GCF}_{A(e,e'np)} \cdot P^{[n]p}_A \cdot T^{NN}_A +$$

$$\sigma^{GCF}_{A(e,e'pn)} \cdot P^{p[n]}_{A} \cdot T^{NN}_{A},$$

$$\sigma^{Exp}_{A(e,e'p)} = \left(\sigma^{GCF}_{A(e,e'pp)} + \sigma^{GCF}_{A(e,e'pn)}\right) \cdot P^{pp}_{A} \cdot T^{N}_{A} + \sigma^{GCF}_{A(e,e'np)} \cdot P^{[n]p}_{A} \cdot T^{N}_{A} + \sigma^{GCF}_{A(e,e'nn)} \cdot P^{[n]n}_{A} \cdot T^{N}_{A},$$

where $\sigma^{GCF}_{X}$ are the GCF simulated events for process X without FSI or SCX, and the $P_A$ and $T_A$ factors are multiplied to the event weights to account for SCX and transparency probabilities, respectively. We note that the $P_A$ and $T_A$ factors do not impact the kinematics of the calculated events.
$T^{NN}_{A}$ refers to the transparency for both the leading and recoil nucleons being emitted simultaneously, while $T^{N}_{A}$ refers to the transparency for the leading nucleon independent of the recoil nucleon. We assume that the transparencies for protons and neutrons are the same, and therefore independent of SCX.
As SCX probabilities are different for protons and neutrons and high and low momentum, the *NN* superscript notation in the P factor mark the exact process being considered, such that particle with (without) square brackets are the ones that undergo (do not undergo) SCX. For example $P^{[p]p}_{A}$ is the probability that a leading proton in a pp pair undergoes SCX, $P^{p[p]}_{A}$ is this probability for the recoil proton and $P^{pp}_{A} = 1 - P^{[p]p}_{A} - P^{p[p]}_{A}$ is the probability that no proton undergoes SCX. As can be seen, SCX change final state neutrons to protons and vice versa. We note that we neglect cases where more than one particle undergoes SCX as these have negligible probability.
The values used for these probabilities are listed in Supplementary Information Table I.

**Table 1 | Single Charge Exchange and Transparency Probabilities:** calculated for $^{12}$C.

| $P^{pp}$ | $P^{[p]p}$ | $P^{p[p]}$ | $P^{[pp]}$ | $P^{p[n]}$ | $P^{[p]n}$ | $P^{np}$ | $P^{[n]p}$ | $P^{n[p]}$ | $P^{[np]}$ | $P^{n[n]}$ | $T^{N}$ | $T^{NN}$ |
|---|---|---|---|---|---|---|---|---|---|---|---|---|
| 90.8% ± 0.6% | 4.1% ± 0.3% | 4.8% ± 0.3% | 0.3% ± 0.02% | 4.1% ± 0.3% | 3.5% ± 0.2% | 92.2% ± 0.5% | 3.5% ± 0.2% | 4.1% ± 0.3% | 0.2% ± 0.01% | 4.8% ± 0.3% | 53% ± 5% | 44% ± 4% |

**Estimate of relativistic effects using the nuclear light-cone formalism:**
The nuclear LC formalism allows accounting for relativistic effects in the two-body nuclear wave function. Its derivation is detailed in Ref. [9, 51, 52]. Here, we review its incorporation into the GCF model, the results of which are shown in the main text to assess the possible impact of relativistic effects.
In the LC formalism, standard momentum vectors are replaced by ($\alpha$, $p_\perp$) where $p_\perp$ is the component of the momentum vector transverse to **q** and $\alpha \equiv \frac{\sqrt{m_N^2+p^2}-p_{||}}{m_A/A}$. Using these notations, the LC equivalent for the GCF universal functions are given by [9]:

$$\text{Eq. 9} \qquad \rho^{\alpha}_{NN}(\alpha_{rel}, p_{rel,\perp}) = \frac{|\tilde{\varphi}^{\alpha}_{NN}(k)|^2}{2-\alpha_{rel}}\sqrt{m_N^2+k^2},$$

where

$$\text{Eq. 10} \qquad k^2 \equiv \frac{m_N^2+k_\perp^2}{\alpha_{rel}(2-\alpha_{rel})} - m_N^2$$

is the effective pair relative momentum probed relativistically, $k_\perp \equiv \frac{\alpha_i p_{recoil,\perp} - \alpha_{recoil} p_{i,\perp}}{\alpha_i + \alpha_{recoil}}$, and $\tilde{\varphi}^{\alpha}_{NN}$ are the non-relativistic universal functions defined previously, $p_{rel,\perp} \equiv \frac{1}{2}(p_{i,\perp} - p_{recoil,\perp})$, and $a_{rel} \equiv \frac{2\alpha_{recoil}}{\alpha_i + \alpha_{recoil}}$. See Supplementary Information Fig. 1 below for details. Similarly, the SRC pair c.m. distribution is given by:

$$\text{Eq. 11} \qquad \rho_{CM}(\alpha_{CM}, p_{CM,\perp}) = \frac{(m_A/A)\alpha_{CM}}{(2\pi\sigma_{CM})^{3/2}} e^{-\frac{\left(\frac{m_A}{A}\right)^2(2-\alpha_{CM})^2+p_{CM,\perp}^2}{2\sigma_{CM}^2}},$$

Where $p_{CM,\perp} \equiv \boldsymbol{p}_{i,\perp} + \boldsymbol{p}_{recoil,\perp}$, and $\alpha_{cm} \equiv \alpha_i + \alpha_{recoil}$. Performing all appropriate variable substitutions and computing the relevant Jacobian, one finds that the LC equivalent of Eq. 6, i.e.:

Eq. 12
$$\frac{d^8\sigma}{dQ^2\, dx_B\, d\phi_e\, d^3\boldsymbol{P}_{CM}\, d\Omega_{recoil}} =$$

$$\frac{\sigma_{eN}}{32\pi^4\alpha_i}\frac{\alpha_{A-2}}{\alpha_{CM}}\frac{\rho_{CM}(\alpha_{CM}, p_{CM,\perp})}{\epsilon_{A-2}}\frac{\epsilon_N \boldsymbol{p}^2_{recoil}\ \Theta\left(\boldsymbol{p}_{rel} - p^{min}_{rel}\right)}{|\epsilon_{recoil}(\boldsymbol{p}_{recoil} - \boldsymbol{Z}cos\theta_{z,recoil}) + \epsilon_N \boldsymbol{p}_{recoil}|}\frac{\omega}{2E_{beam}E_e x_B}\frac{\sqrt{m_N^2 + k^2}}{2 - \alpha_{rel}}\sum_\alpha C_\alpha |\tilde{\varphi}^\alpha_{NN}(k)|^2,$$

Which can be obtained by simply substituting into the right hand side of Eq. 6: $|\tilde{\varphi}^\alpha_{NN}(\boldsymbol{p}_{rel})|^2 \to \rho^\alpha_{NN}(\alpha, p_\perp)$ and $n_{CM}(\boldsymbol{P}_{CM}) \to \rho_{CM}(\alpha_{CM}, p_{CM,\perp})$ and by inserting the necessary Jacobian factors resulting from the transformation from **p** to $(\alpha, p_\perp)$. Following these modifications, the rest of the LC calculation follows exactly the non-relativistic description above.

The importance of relativistic effects at high-momenta can be seen by considering the simple case of scattering off a forward vs. backward going nucleon in the deuteron. Without accounting for relativistic effects, in the forward scattering case (i.e. recoil nucleon at 180° to **q**) the maximally allowed momenta of the recoil nucleon equals (3/4) $m_N$, while in the backward scattering case (i.e. recoil nucleon at 0° to **q**) there is no kinematical restriction on the momenta of the recoil nucleon. The LC formalism presented above removes this asymmetry [9].

As mentioned in the main text, while studied in detail [9, 51, 52] and used by previous works to analyze proton induced knockout reactions [15], the LC prescription used here to account for relativistic effects is approximate and model dependent. The accuracy of Eq. 9 was previously studied by Ref. [53] using simple covariant models for which the four-dimensional solution of the Bethe-Salpeter wave function can be obtained, to find that for these models Eq. 9 requires corrections on the scale of 5% - 10% for the kinematics of the current experiment. This is encouraging; however, these estimations are based on simple models and should be extended in the future for more realistic interactions.

**Tests of Contributions from Non-QE Reaction Mechanisms:**

Non-QE reaction mechanisms, beyond those accounted for by the SCX and transparency corrections, such as small-angle leading-nucleon rescattering, can modify the measured kinematics and therefore interfere with the interpretation of the data. By changing the leading-nucleon momentum, rescattering can cause events with high missing-momentum that originate from interactions with low initial-momentum nucleons. We performed several experimental tests of these effects.

(1) Rescattering should increase with atomic mass, but the properties of SRCs should be very similar for different nuclei [9, 7, 8, 44]. Therefore we examined the nuclear mass (A) dependence of the data, for A = 12 (C), 27 (Al), 56 (Fe), and 208 (Pb). See Extended Data Figs. 2 and 3 for the missing momentum, $x_B$, $Q^2$ and angle between missing-momentum and momentum transfer (q) of the measured A(e,e'p) and A(e,e'pp) event yields, as well as for the missing momentum dependence of the measured A(e,e'pp) / A(e,e'p) yield ratio. In all cases the data for the different nuclei are very similar, indicating that A-dependent effects are small.

(2) Leading-nucleon rescattering would give a peak in the $\theta_{p_{miss}q}$ (the angle between the momentum transfer and the missing momentum) distribution at 110° (non-relativistically, that peak would be at 90°) [8, 19]. Extended Data Fig. 3 shows the $\theta_{p_{miss}q}$ distribution for all four nuclei and the GCF $^{12}$C calculations. There is no peak in either the data or the calculation at the expected rescattering maximum. In addition, the $\theta_{p_{miss}q}$ distributions are similar for all nuclei, whereas rescattering should increase with A. These are further indications that rescattering is small for this data.

(3) Light-cone momentum densities are sensitive to longitudinal momentum components relative to the momentum transfer [17, 19]. Calculations show that while nucleons that undergo rescattering change both their energy and momenta, at large-$Q^2$ and anti-parallel kinematics the difference between the nucleon energy and its momentum component along the q vector direction, which is proportional to its light-cone momentum, is approximately conserved [19]. Therefore, we expect for light-cone momentum distributions to be well reproduced by the GCF calculation.

The light-cone momentum fraction carried by the interacting nucleon is defined as: $\alpha_{miss} \equiv \alpha_N - \alpha_q$, where $\alpha_N \equiv \frac{E_N - p_N^{\hat{q}}}{m_A/A}$ and $\alpha_q \equiv \frac{\omega - q}{m_A/A}$. Extended Data Fig. 7 shows the distribution of $\alpha_{miss}$ for (e,e'p) and (e,e'pp) reactions for both data and GCF calculations. For completeness, it also shows the light-cone pair CM momentum distribution $\alpha_{c.m.} \equiv \alpha_N + \alpha_{recoil}$ (where $\alpha_{recoil} \equiv \frac{E_{recoil} - p_{recoil}^{\hat{q}}}{m_A/A}$) for the (e,e'pp) reaction for both data and GCF calculations. $\alpha_{miss}$ ranges from 1.2 to about 1.6, spanning the expected range for 2N-SRC pairs dominance [17]. $\alpha_{c.m.}$ is centered around the expected value of 2. As expected, all light-cone momentum distributions show overall good agreement between the data and calculations.

(4) As the data and calculations shown in Extended Data Fig. 3 are integrated over missing momentum, they are dominated by low missing-momentum, potentially masking issues at high missing momentum. To address this, Supplementary Information Fig. 2 show the distribution of the components of the missing momentum in the direction longitudinal and transverse to the momentum-transfer q, in bins of missing momentum for the (e,e'p) and (e,e'pp) reactions for both data and GCF calculations. In all missing-momentum bins, the data and simulation show good agreement, and the kinematics are predominantly anti-parallel with the longitudinal component being larger than the transverse.

To supplement the light-cone momentum density discussed in point (3) above (which is sensitive to the longitudinal component of the missing momentum), we also note that we do not see any enhancement of the transverse component of the missing momentum, as compared with the GCF expectation. Such enhancement is a typical signature of small-angle elastic scattering, which is not included in the GCF calculation but could be present in the data. The agreement of the data with the GCF calculation suggest that such scattering does not contribute significantly to the data.

(5) Lastly, we note the different missing momentum dependence of the measured and simulated (e,e'p) and (e,e'pp) event yields. As can be seen in Fig. 3, while the (e,e'p) distribution falls over two orders of magnitude, the (e,e'pp) distribution is much flatter and only varies over one order of magnitude. If both distributions were driven by rescattering of low initial-momentum nucleons as they exit the nucleus, leading to the emission of a recoil nucleon and formation of large missing momentum, both distributions should have similar missing-momentum dependence.

We note that all conclusions mentioned above hold true also when relativistic corrections are introduced to the calculations.

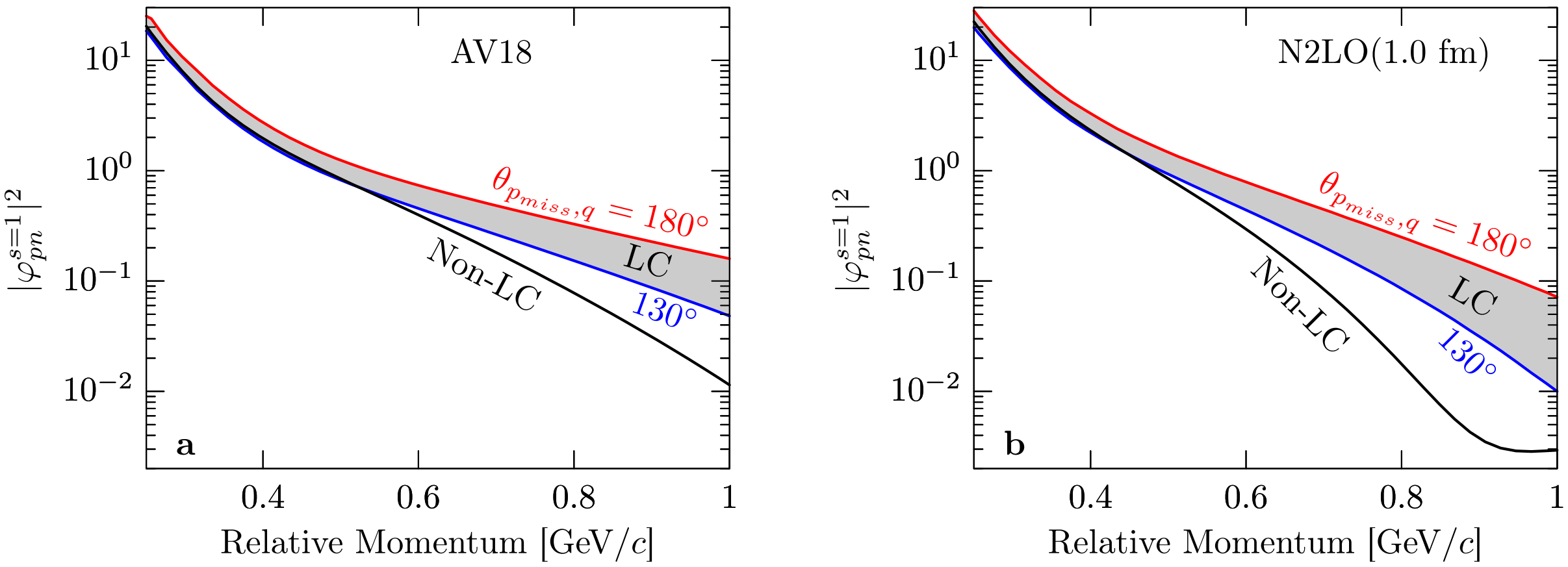

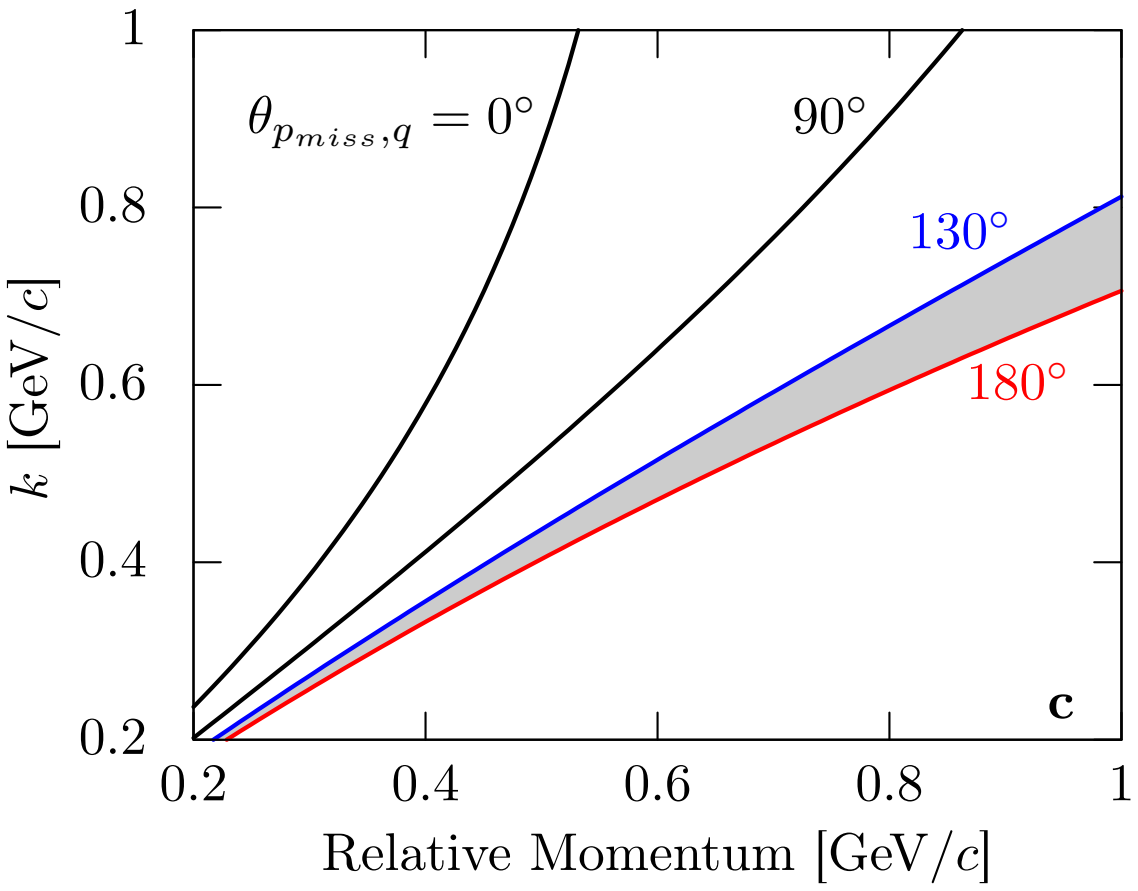

**Supplementary Fig. 1 | Relative Momenta in Light-Cone Formalism.** Bottom Left: the LC relative momentum variable, *k,* vs. the non-relativistic relative momentum for several different angles between $\boldsymbol{p_{miss}}$ and $\boldsymbol{q}$. The kinematics from this measurement are shown by the gray band. Top: the effective relative momentum distribution calculated for the AV18 (a) and N2LO(1.0) (b) potentials in non-LC (black line) and LC (gray band) approaches for the range of $\theta_{p_{miss},q}$ angles as in the data. The LC distribution is larger in the high-momentum tail region because $k$ is smaller than the relative momentum.

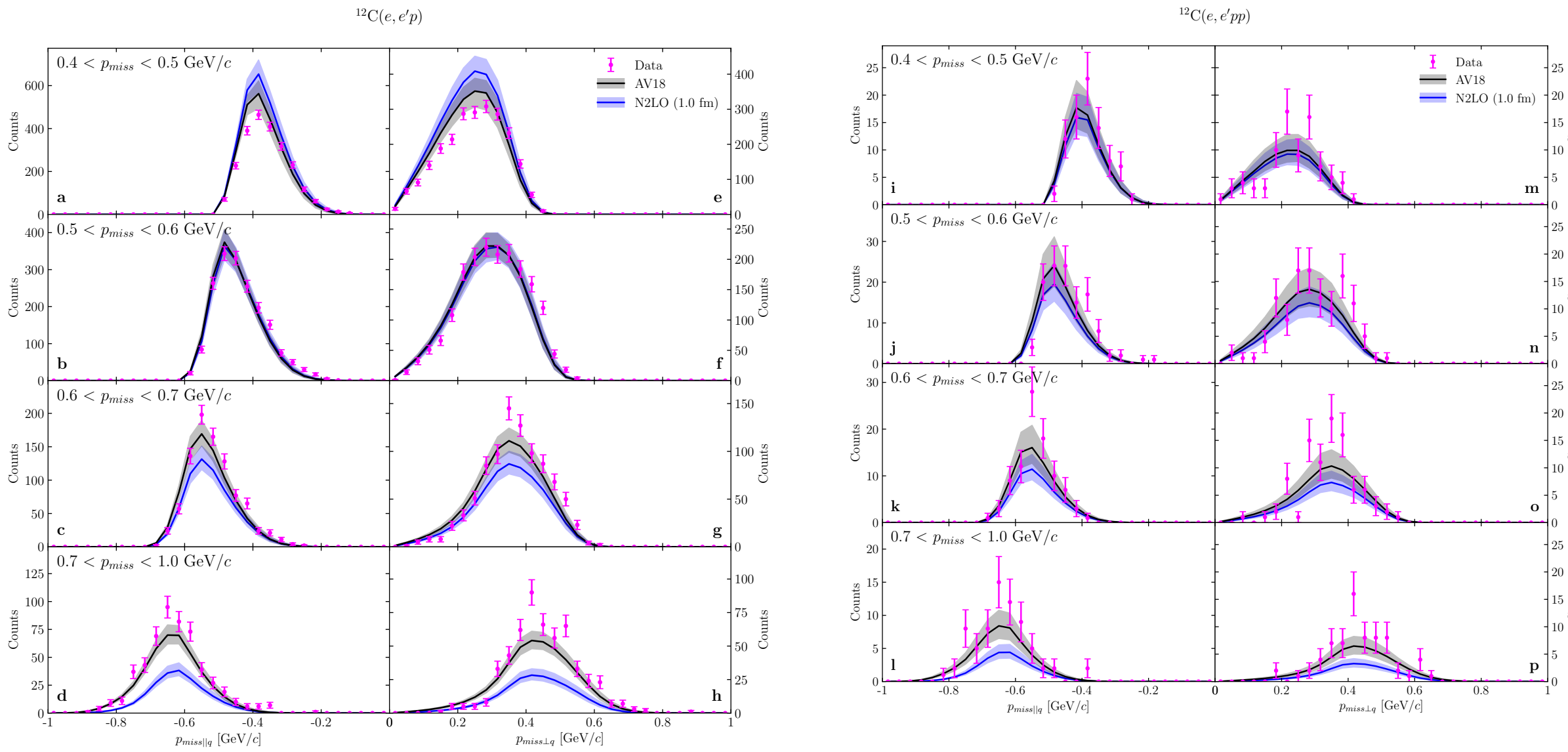

**Supplementary Fig. 2 | Missing momentum components for $^{12}$C(e,e'p) and $^{12}$C(e,e'pp) events:** Distributions of the components of the $^{12}$C(e,e'p) (a - h) and $^{12}$C(e,e'pp) (i - p) missing momentum parallel (a – d, i - l) and perpendicular (e - h, m - p) to the momentum transfer q for different missing-momentum vector magnitudes. Colored bands show GCF calculations using different *NN* interaction models The width of the band and the data error bars show the model systematic uncertainties and data statistical uncertainties, respectively, each at the 1σ or 68% confidence level.

## References:

[1] "Quantum Monte Carlo methods for nuclear physics", J. Carlson, S. Gandolfi, F. Pederiva, S. C. Pieper, R. Schiavilla, K. E. Schmidt, and R. B. Wiringa, Rev. Mod. Phys. **87**, 1067 (2015).

[2] "Modern theory of nuclear forces", E. Epelbaum, H.W. Hammer and U.G. Meißner, Rev. Mod. Phys. **81**, 1773 (2009).

[3] "The bonn meson-exchange model for the nucleon-nucleon interaction", R. Machleidt, K. Holinde, and C. Elster, Phys. Rept. **149**, 1-89 (1987).

[4] "Accurate nucleon-nucleon potential with charge-independence breaking", R. B. Wiringa, V. G. J. Stoks, and R. Schiavilla, Phys. Rev. C **51**, 38 (1995).

[5] "Quantum Monte Carlo Calculations with Chiral Effective Field Theory Interactions", A. Gezerlis, I. Tews, E. Epelbaum, S. Gandolfi, K. Hebeler, A. Nogga, and A. Schwenk, Phys. Rev. Lett. **111**, 032501 (2013).

[6] "Neutron Star Observations: Prognosis for Equation of State Constraints" J.M. Lattimer and M. Prakash, Phys. Rept. **442**, 109-165 (2007).

[7] "Nucleon-Nucleon Correlations, Short-lived Excitations, and the Quarks Within", O. Hen, G.A. Miller, E. Piasetzky, and L. B. Weinstein, Rev. Mod. Phys. **89**, 045002 (2017).

[8] "In-medium short-range dynamics of nucleons: Recent theoretical and experimental advances", C. Ciofi degli Atti, Phys. Rep. **590**, 1-85 (2015).

[9] "High-Energy Phenomena, Short Range Nuclear Structure and QCD", L. Frankfurt, and M. Strikman, Phys. Rep. **76**, 215-347 (1981).

[10] "Probing Cold Dense Nuclear Matter", R. Subedi et al., Science **320**, 1476-1478 (2008).

[11] "Approaching the nucleon-nucleon short-range repulsive core via the 4He(e,e'pN) triple coincidence reaction", I. Korover et al., Phys. Rev. Lett. **113**, 022501 (2014).

[12] "Momentum Sharing in Imbalanced Fermi Systems", O. Hen et al. (CLAS Collaboration), Science **346**, 614-617 (2014).

[13] "Direct Observation of Proton-Neutron Short-Range Correlation Dominance in Heavy Nuclei", M. Duer et al. (CLAS Collaboration), Phys. Rev. Lett. **122**, 172502 (2019).

[14] "Nucleon knockout by intermediate-energy electrons", J.J. Kelly, Adv. Nucl. Phys. **23**, 75-294 (1996).

[15] "Evidence for the strong dominance of proton-neutron correlations in nuclei", E. Piasetzky, M. Sargsian, L. Frankfurt, M. Strikman, and J.W. Watson, Phys. Rev. Lett. **97**, 162504 (2006).

[16] "Center of Mass Motion of Short-Range Correlated Nucleon Pairs studied via the A(e,e'pp) Reaction", E. O. Cohen et al. (CLAS Collaboration), Phys. Rev. Lett. **121**, 092501 (2018).

[17] "Recent observation of short range nucleon correlations in nuclei and their implications for the structure of nuclei and neutron stars", L. Frankfurt, M. M. Sargsian, and M. Strikman, Int. J. Mod. Phys. A **23**, 2991-3055 (2008).

[18] "Final-state interactions in two-nucleon knockout reactions", C. Colle, W. Cosyn, and J. Ryckebusch, Phys. Rev. C **93**, 034608 (2016).

[19] "Selected Topics in High Energy Semi-Exclusive Electro-Nuclear Reactions", M. M. Sargsian, Int. J. Mod. Phys. E **10**, 405-458 (2001).

[20] "The CEBAF large acceptance spectrometer (CLAS)", B. A. Mecking et al., Nucl. Inst. Meth. A **503**, 513-553 (2003).

[21] "Off-shell electron-nucleon cross sections: The impulse approximation", T. De Forest, Nucl. Phys. A **392**, 232-248 (1983).

[22] "Energy and momentum dependence of nuclear short-range correlations - Spectral function, exclusive scattering experiments and the contact formalism", R. Weiss, I. Korover, E. Piasetzky, O. Hen, and N. Barnea, Phys. Lett. B **791**, 242-248 (2019).

[23] "Generalized nuclear contacts and momentum distributions", R. Weiss, B. Bazak, and N. Barnea, Phys. Rev. C **92**, 054311 (2015).

[24] "The Nuclear Contacts and Short-Range Correlations in Nuclei", R. Weiss, R. Cruz-Torres, N. Barnea, E. Piasetzky, and O. Hen, Phys. Lett. B **780**, 211-215 (2018).

[25] "Scale and Scheme Independence and Position-Momentum Equivalence of Nuclear Short-Range Correlations" R. Cruz-Torres, D. Lonardoni, R. Weiss, N. Barnea, D.W. Higinbotham, E. Piasetzky, A. Schmidt, L.B. Weinstein, R.B. Wiringa, and O. Hen, arXiv: 1907.03658 (2019).

[26] "Weinberg eigenvalues for chiral nucleon-nucleon interactions", J. Hoppe, C. Drischler, R. J. Furnstahl, K. Hebeler, and A. Schwenk Phys. Rev. C 96, 054002 (2017).

[27] "Two nucleon correlations and the structure of the nucleon spectral function at high values of momentum and removal energy", C. Ciofi degli Atti, S. Simula, L.L. Frankfurt, and M.I. Strikman, Phys. Rev. C **44**, R7-R11 (1991).

[28] "The EMC Effect", P.R. Norton, Rept. Prog. Phys. **66**, 1253-1297 (2003).

[29] "Global study of nuclear structure functions", S. A. Kulagin and R. Petti, Nucl. Phys. A **765**, 126-187 (2006).

[30] “Modified Structure of Protons and Neutrons in Correlated Pairs”, B. Schmookler et al. (CLAS Collaboration), Nature **566**, 354-358 (2019).
[31] “A Double Target System for Precision Measurements of Nuclear Medium Effects”, H. Hakobyan et al., Nucl. Inst. and Meth. A **592**, 218-223 (2008).
[32] “Short range correlations and the isospin dependence of nuclear correlation functions”, R. Cruz-Torres, A. Schmidt, G.A. Miller, L.B. Weinstein, N. Barnea, R. Weiss, E. Piasetzky, and O. Hen, Phys. Lett. B 785, 304-308 (2018).
[33] “Short-range correlations and the charge density”, R. Weiss, A. Schmidt. G.A. Miller, and N. Barnea, Phys. Lett. B **790**, 484-489 (2019).
[34] “Exclusive electrodisintegration of 3He at high *Q*2 II. Decay function formalism” M.M. Sargsian, T. V. Abrahamyan, M. I. Strikman, and L. L. Frankfurt, Phys. Rev. C 71, 044615 (2005).
[35] “Feynman graphs and generalized eikonal approach to high energy knock-out processes”, L.L. Frankfurt, M.M. Sargsian, and M.I. Strikman, Phys. Rev. C **56**, 1124 (1997).
[36] “Quasielastic 3He(e,e’p)2H Reaction at *Q*2 = 1.5 GeV2 for Recoil Momenta up to 1 GeV/c”, M. Rvachev et al. Phys. Rev. Lett. **94**, 192302 (2005).
[37] “Measurement of the 3He(e,e’p)pn Reaction at High Missing Energies and Momenta”, F. Benmokhtar et al. Phys. Rev. Lett. **94**, 082305 (2005).
[38] “Experimental Study of Exclusive $^2$H(e,e’p)n Reaction Mechanisms”, K. S. Egiyan et al. (CLAS Collaboration), Phys. Rev. Lett. **98**, 262502 (2007).
[39] “Probing the High Momentum Component of the Deuteron at High $Q^2$”, W.U. Boeglin et al. Phys. Rev. Lett. **107**, 262501 (2011).
[40] “Color transparency: past, present and future”, D. Dutta, K. Hafidi, and M. Strikman, Prog. Part. Nucl. Phys. **69**, 1-27 (2013).
[41] “Measurement of Transparency Ratios for Protons from Short-Range Correlated Pairs”, O. Hen et al. (CLAS Collaboration), Phys. Lett. B **722**, 63-68 (2013).
[42] “Measurement of Nuclear Transparency Ratios for Protons and Neutrons”, M. Duer, et al. (CLAS Collaboration), Phys. Lett. B **797**, 134792 (2019).
[43] “Extracting the mass dependence and quantum numbers of short-range correlated pairs from A(e,e’p) and A(e,e’pp) scattering”, C. Colle et al., Phys. Rev. C **92**, 024604 (2015).
[44] “Universality of many-body two-nucleon momentum distributions: Correlated nucleon spectral function of complex nuclei”, C. Ciofi degli Atti and H. Morita, Phys. Rev. C **96**, 064317 (2017).
[45] “n-p short range correlations from (p,2p + n) measurements”, A. Tang et al (EVA Collaboration), Phys. Rev. Lett. **90**, 042301 (2003).
[46] “Investigation of proton-proton short-range correlations via the 12C(e,e’pp) reaction”, R. Shneor et al. (Jefferson Lab Hall A Collaboration), Phys. Rev. Lett. **99**, 072501 (2007).
[47] “Single- and two-nucleon momentum distributions for local chiral interactions”, D. Lonardoni, S. Gandolfi, X. B. Wang, and J. Carlson, Phys. Rev. C **98**, 014322 (2018).
[48] “Nucleon and nucleon-pair momentum distributions in A ≤ 12 nuclei”, R.B. Wiringa, R. Schiavilla, S.C. Pieper, and J. Carlson, Phys. Rev. C **89**, 024305 (2014).
[49] “Realistic model of the nucleon spectral function in few and many nucleon systems”, C. Ciofi degli Atti and S. Simula, Phys. Rev. C **53**, 1689 (1996).
[50] “Factorization of exclusive electron-induced two-nucleon knockout”, C. Colle, W. Cosyn, J. Ryckebusch, and M. Vanhalst, Phys. Rev. C **89**, 024603 (2014).
[51] “Short range correlations in nuclei as seen in hard nuclear reactions and light cone dynamics”, L. Frankfurt and M. Strikman, Modern topics in electron scattering, B. Frois (ed.), I. Sick (ed.), 645-694, (1992).
[52] “Multinucleon short-range correlation model for nuclear spectral functions: Theoretical framework”, O. Artiles and M. Sargsian, Phys. Rev. C **94**, 064318 (2016).
[53] “Relation between equal-time and light-front wave functions”, G.A. Miller and B.C. Tiburzi, Phys. Rev. C **81**, 035201 (2010).
[54] “Radiative corrections for (e,e’p) reactions at GeV energies”, R. Ent, B.W. Filippone, N.C.R. Makins, R.G. Milner, T.G. O’Neill, and D.A. Wasson, Phys. Rev. C **64**, 054610 (2001).
[55] “Scale dependence of deuteron electrodisintegration”, S. N. More, S. K. Bogner, and R. J. Furnstahl, Phys. Rev. C **96**, 054004 (2017).
[56] “Quantum Monte Carlo calculations of light nuclei with local chiral two- and three-nucleon interactions”, J. E. Lynn, I. Tews, J. Carlson, S. Gandolfi, A. Gezerlis, K. E. Schmidt, and A. Schwenk, Phys. Rev. C **96**, 054007 (2017)